\documentclass[aps,prl,nofootinbib,twocolumn,superscriptaddress,floatfix,notitlepage]{revtex4-1}
\pdfoutput=1

\usepackage{graphicx}
\usepackage{amsmath,amsfonts,amssymb}
\usepackage{color}
\usepackage[breaklinks,colorlinks,urlcolor=blue,citecolor=blue,linkcolor=magenta]{hyperref}
\usepackage{verbatim}
\usepackage{enumitem}
\usepackage{hyperref}
\usepackage{natbib}
\usepackage{aas_macros}

\newcommand{\be}{\begin{equation}} 
\newcommand{\ee}{\end{equation}}
\newcommand{\bea}{\begin{equation}\begin{aligned}} 
\newcommand{\eea}{\end{aligned}\end{equation}}
\newcommand{\td}{{\rm d}}
\newcommand{\cM}{\mathcal{M}}
\newcommand{\Msun}{M_\odot}

\newcommand{\yr}{{\rm yr}}
\newcommand{\ie}{{{i.e.}}}
\newcommand{\eg}{{{e.g.}}}

\def\lsim{\mathrel{\raise.3ex\hbox{$<$\kern-.75em\lower1ex\hbox{$\sim$}}}}
\def\gsim{\mathrel{\raise.3ex\hbox{$>$\kern-.75em\lower1ex\hbox{$\sim$}}}}

\begin{document}

\title{Gravitational Waves from SMBH Binaries in Light of the NANOGrav 15-Year Data}

\author{John Ellis}
\email{john.ellis@cern.ch}
\affiliation{Keemilise ja Bioloogilise F\"u\"usika Instituut, R\"avala pst. 10, 10143 Tallinn, Estonia}
\affiliation{Physics Department, King’s College London, Strand, London, WC2R 2LS, United Kingdom}
\affiliation{Theoretical Physics Department, CERN, CH 1211 Geneva, Switzerland}

\author{Malcolm Fairbairn}
\email{malcolm.fairbairn@kcl.ac.uk}
\affiliation{Physics Department, King’s College London, Strand, London, WC2R 2LS, United Kingdom}

\author{Gert H\"utsi}
\email{gert.hutsi@kbfi.ee}
\affiliation{Keemilise ja Bioloogilise F\"u\"usika Instituut, R\"avala pst. 10, 10143 Tallinn, Estonia}

\author{Juhan Raidal}
\email{juhan.raidal@student.manchester.ac.uk}
\affiliation{Keemilise ja Bioloogilise F\"u\"usika Instituut, R\"avala pst. 10, 10143 Tallinn, Estonia}

\author{Juan Urrutia}
\email{juan.urrutia@kbfi.ee}
\affiliation{Keemilise ja Bioloogilise F\"u\"usika Instituut, R\"avala pst. 10, 10143 Tallinn, Estonia}
\affiliation{Departament of Cybernetics, Tallinn University of Technology, Akadeemia tee 21, 12618 Tallinn, Estonia}

\author{Ville Vaskonen}
\email{ville.vaskonen@pd.infn.it}
\affiliation{Keemilise ja Bioloogilise F\"u\"usika Instituut, R\"avala pst. 10, 10143 Tallinn, Estonia}
\affiliation{Dipartimento di Fisica e Astronomia, Universit\`a degli Studi di Padova, Via Marzolo 8, 35131 Padova, Italy}
\affiliation{Istituto Nazionale di Fisica Nucleare, Sezione di Padova, Via Marzolo 8, 35131 Padova, Italy}

\author{Hardi Veerm\"ae}
\email{hardi.veermae@cern.ch}
\affiliation{Keemilise ja Bioloogilise F\"u\"usika Instituut, R\"avala pst. 10, 10143 Tallinn, Estonia}

\begin{abstract}
The NANOGrav and other Pulsar Timing Arrays (PTAs) have recently announced evidence for nHz gravitational waves (GWs) that may originate from supermassive black holes (SMBH) binaries. The spectral index of the GW signal differs from that predicted for binary evolution by GW emission alone, and we show that environmental effects such as dynamical friction with gas, stars and dark matter improve the consistency of the SMBH binary model with the PTA data. We comment on the possible implications of environmental effects for PTA observations of fluctuations in the GW frequency spectrum and measurements of GWs at higher frequencies. 
\\~~\\
KCL-PH-TH/2023-37, CERN-TH-2023-120, AION-REPORT/2023-06
\end{abstract}

\maketitle

%%%%%%%%%%%%%%%%%%%%%%%%%%%%%%%%%%%%%%%%%%%
\noindent\textbf{Introduction --} 
%%%%%%%%%%%%%%%%%%%%%%%%%%%%%%%%%%%%%%%%%%%
Several ongoing Pulsar Timing Array (PTA) projects -- the North American Nanohertz Observatory for Gravitational Waves (NANOGrav), the European PTA,  the Parkes PTA and the Chinese PTA -- have recently released their latest data~\cite{NANOGrav:2023hde,EPTA:2023sfo,Zic:2023gta,Xu:2023wog}.
Most importantly, the NANOGrav Collaboration has found evidence~\cite{NANOGrav:2023gor} in NANOGrav 15-year (NG15) data for the Hellings-Downs quadrupolar correlation~\cite{Hellings:1983fr} (see also~\cite{Antoniadis:2023ott,Reardon:2023gzh,Xu:2023wog}) - a key characteristic of gravitational waves (GWs) - in the common-spectrum process observed previously by them~\cite{Arzoumanian:2020vkk} and other PTAs~\cite{Goncharov:2021oub,Chen:2021rqp,Antoniadis:2022pcn}. This is a landmark in GW astronomy, of significance comparable to the first indirect detection of GWs emitted by binary pulsars~\cite{Hulse:1974eb} and the first direct observation of GW emissions from stellar-mass black holes (BH) binaries~\cite{LIGOScientific:2016aoc}. This discovery of a nHz stochastic GW background opens a new window on astrophysical processes that were previously unobserved. We assume here that the origin of the PTA signal is a population of Supermassive Black Hole (SMBH) binaries (see also~\cite{NANOGrav:2023hfp,Antoniadis:2023xlr}), and explore the astrophysical implications of this possibility.\footnote{Numerous interpretations of the NANOGrav 12.5-year data~\cite{Arzoumanian:2020vkk} based on cosmological sources have been put forward, including cosmic strings and domain walls~\cite{Ellis:2020ena,Datta:2020bht,Samanta:2020cdk,Buchmuller:2020lbh,Blasi:2020mfx,Buchmuller:2021mbb,Blanco-Pillado:2021ygr,Ferreira:2022zzo,An:2023idh,Qiu:2023wbs,Zeng:2023jut}, first-order phase transitions~\cite{NANOGrav:2021flc,Xue:2021gyq,Nakai:2020oit,DiBari:2021dri,Sakharov:2021dim,Li:2021qer,Ashoorioon:2022raz,Benetti:2021uea,Barir:2022kzo,Hindmarsh:2022awe}, and primordial fluctuations~\cite{Vaskonen:2020lbd,DeLuca:2020agl,Bhaumik:2020dor,Inomata:2020xad,Kohri:2020qqd,Domenech:2020ers,Vagnozzi:2020gtf,Namba:2020kij,Sugiyama:2020roc,Zhou:2020kkf,Lin:2021vwc,Rezazadeh:2021clf,Kawasaki:2021ycf,Ahmed:2021ucx,Yi:2022ymw,Yi:2022anu,Dandoy:2023jot,Zhao:2023xnh,Ferrante:2023bgz,Cai:2023uhc}. Many of these models with a spectral slope $\gamma > 13/3$ are disfavoured by the current data~\cite{NANOGrav:2023hvm}, for instance, cosmic strings or primordial black hole seeds for SMBHs.} 

Accretion around SMBHs drives Active Galactic Nuclei (AGNs), and is thought to play a role in the formation of SMBH binaries~\cite{Mayer:2013jja,Foreman:2008th}. At an early stage, the binary evolution is driven by dynamical friction with gas, stars, and dark matter~\cite{Begelman:1980vb}. However, a prerequisite for SMBH mergers is the formation of tightly bound binaries that can radiate GWs efficiently. It is not yet understood how this stage of SMBH binary evolution is reached, an issue commonly known as the ``final parsec problem''~\cite{Begelman:1980vb}. If the binaries can overcome it, an important source of evolution at subparsec scales is energy loss via GW emission until it is the main driver of the evolution close to merging. SMBH binaries, driven only by GW emission, are naively expected to produce an almost flat background with a spectral index $\gamma = 13/3$~\cite{Phinney:2001di}. The recent NG15 data~\cite{NANOGrav:2023gor} on GWs, although compatible with this value at the 99\% CL, prefer a lower value of $\gamma$. This hints that the GWs may be emitted by SMBH binaries that are experiencing additional energy loss via environmental effects, which might be related to whatever dynamics overcomes the final parsec problem.

In this paper, we follow~\cite{Ellis:2023owy} in using the Extended Press-Schechter (EPS) formalism to estimate the SMBH binary formation and merger rates. {We go beyond~\cite{Ellis:2023owy} by including scatter in the SMBH-halo mass relation and by} modifying the evolution of the SMBH binaries to include energy loss via environmental effects as well as GWs.
{
Our analysis improves existing SMBH interpretations~\cite{NANOGrav:2023hfp} of the NG15 data by accurately modeling the long-tailed statistical fluctuations in the GW spectrum without relying on Gaussian approximations.
}
We find that the best fit to the NG15 data~\cite{NANOGrav:2023gor} is given by a phenomenological model of SMBH binary evolution that includes environmental energy loss, with a log-likelihood difference of $10$ compared to purely GW-driven evolution, and hence is favored by more than $2 \sigma$. Within this framework, we find that the efficiency for the production of SMBH mergers, $p_{\rm BH}$, is probably quite high and close to the maximal allowed value. These results provide important constraints on astrophysical scenarios for the interlinked dynamics of SMBHs and their host galaxies, and suggest promising rates for observing mergers of lower-mass BHs in detectors sensitive to GWs with higher frequencies.

%%%%%%%%%%%%%%%%%%%%%%%%%%%%%%%%%%%%%%%%%%%
\vspace{5pt}\noindent\textbf{SMBH background with environmental effects --}
%%%%%%%%%%%%%%%%%%%%%%%%%%%%%%%%%%%%%%%%%%%
The mean of the GW energy density spectrum generated by a population of SMBH binaries can be estimated as~\cite{Phinney:2001di}
\be \label{eq:OmegaGW}
    \Omega_{\rm GW}(f) \equiv \frac{1}{\rho_{\rm c}} \frac{\td \rho_{\rm GW}}{\td \ln f}
    = \frac{1}{\rho_{\rm c}}\int \text{d}\lambda \,\frac{1+z}{4\pi D_L^2} \frac{\td E_{\rm GW}}{\td \ln f_r} \,,
\ee
where $D_L$ denotes the source luminosity distance, $\td E_{\rm GW}/\td \ln f_r$ gives the energy emitted by the source per logarithmic frequency interval and $f_r \equiv (1+z) f$ denotes the frequency in the source frame, \ie, at the time of emission. The differential BH merger rate {in the observer frame} is
\be
    \td \lambda 
    = \td \cM \td \eta \td z \frac{1}{1+z} \frac{\td V_c}{\td z} \frac{\td R_{\rm BH}}{\td \cM \td \eta} \, ,
\ee
where $\cM$ denotes the binary chirp mass, $\eta$ its symmetric mass ratio, $V_c$ the comoving volume, and $R_{\rm BH}$ the comoving BH merger rate density. It is given by 
\bea
    \frac{\td R_{\rm BH}}{\td m_1 \td m_2} =
&   p_{\rm BH}(m_1,m_2,z) \int \td M_1 \td M_2 \, \frac{\td R_h}{\td M_1 \td M_2}\\
&   \times p_{\rm occ}(m_1|M_1,z) p_{\rm occ}(m_2|M_2,z)  \,, 
\eea
where $m_{1,2}$ are the masses of the merging BHs, $M_{1,2}$ are the masses of their host halos, $R_h$ is the halo merger rate that we estimate with the EPS formalism~\cite{Press:1973iz,Bond:1990iw,Lacey:1993iv} and $p_{\rm BH} \le 1$ combines the SMBH occupation fraction in galaxies with the efficiency for the BHs to merge following the merging of their host halos. For simplicity, we assume a constant value for $p_{\rm BH}$, which we treat as a free parameter to be determined by fitting the NG15 data. As the GW background arises from a relatively narrow range of SMBH masses and redshifts, we expect that extending the parametrization by letting $p_{\rm BH}$ vary would have a minor effect on our conclusions. The halo mass-SMBH mass relation is encoded in $p_{\rm occ}$ that is the probability distribution of the SMBH mass.\footnote{Compared to Eq.~(5) of Ref.~\cite{Ellis:2023owy}, the normalization of $p_{\rm occ}$ is absorbed into $p_{\rm BH}$.} We model the spread in the mass relation with a log-normal distribution:
\be
    p_{\rm occ}(m|M,z) = \frac{1}{\sqrt{2\pi}m\sigma}\exp\!\left[-\frac{\ln(m/\bar{m})^2}{2\sigma^2}\right]\,,
\ee
with $\log_{10}(\bar{m}/M_\odot) = 8.95 + 1.4 \log_{10} (M_*/10^{11}M_\odot)$ and $\sigma = 1.1$ following the fit to observations of inactive galaxies in~\cite{Kormendy:2013dxa,2015ApJ...813...82R}, and relate the stellar mass $M_*$ to the halo mass $M_{\rm halo}$ using the fit provided in~\cite{Girelli:2020goz}. The effect of using the fit to AGN and not including the scatter is discussed in the Supplemental Material.  

The GW spectrum from an inspiralling binary is determined by its orbital evolution and, following NANOGrav~\cite{NANOGrav:2023gor} we assume circular binaries. The eccentricity of the orbits (considered, e.g., in~\cite{EPTA:2023xxk}) would affect the GW spectrum by introducing higher harmonics, increasing the total power emitted in GWs and modifying the frequency spectral index~\cite{Enoki:2006kj}. However, big eccentricities $e>0.9$ would lead to an attenuation of the background due to the acceleration of the binary inspiral~\cite{Kelley:2017lek}.

The binary may lose energy through a combination of GW emission and dissipative environmental effects:
\be
    \dot{E} = -\dot{E}_{\rm GW} -\dot{E}_{\rm env}\, ,
\ee
where the dot denotes differentiation with respect to time. This energy loss causes orbital decay and growth in the orbital frequency. 
SMBH binaries are thought to go through several stages as they harden, starting with dynamical friction, followed by hardening due to close stellar encounters with stars populating the so-called `loss-cone' orbits~\cite{Merritt:2013awa}, and finally, dissipation caused by viscous drag due to circumbinary gas disks~\cite{Begelman:1980vb}. 
The crossing of the final parsec is determined mostly by these last two effects, with the latter being the dissipation mechanism that is commonly thought to remain active after the GW emission has become significant, see \eg~\cite{Armitage:2002uu,Macfadyen:2006jx,Tang:2017eiz}, thus leading to faster binary hardening compared to the GW-only case.

The impact of the gas is not fully established, \eg,~\cite{Munoz:2018tnj,Moody:2019nes} argue that accretion from circumbinary disks might have quite the opposite effect -- instead of hardening the binary, it might lead to the expansion of the orbit. More recent work predicts, however, inspiralling for more realistic cooler, thinner disks~\cite{Tiede:2020ldm}, non-zero eccentricities~\cite{DOrazio:2021kob} and unequal masses~\cite{Duffell:2019uuk}. Any slowing of the infall of binaries contributing in the low-frequency bins would worsen the SMBH fit, whereas extra dissipation improves it, as we show below.

The characteristic timescales for the above processes are 
\be
    t_{\rm GW} \equiv |E|/\dot{E}_{\rm GW} = 4 \tau \, ,
    \qquad
    t_{\rm env} \equiv |E|/\dot{E}_{\rm env} \, ,
\ee
where $\tau = (5/256) (\pi f_r)^{-8/3} \cM^{-5/3}$ denotes the coalescence time of the binary assuming GW emission alone. The effective timescale for the binary evolution is then $t_{\rm eff}^{-1}=t_{\rm GW}^{-1}+t_{\rm env}^{-1}$. Since the binding energy of the binary is $E = -\left(\pi f_r \right)^{2/3} \cM^{5/3}/2$, where $f_r$ is the frequency of the emitted GW, it follows that
\be
    \frac{\td \ln f_r}{\td t} = \frac32 t_{\rm GW}^{-1} \left(1 + \frac{t_{\rm GW}}{t_{\rm env}}\right)\, ,
\ee
and thus
\be\label{eq:EGW}
    \frac{\td E_{\rm GW}}{\td \ln f_r} 
    = \frac{\td E_{\rm GW}}{\td t} \frac{\td t}{\td \ln f_r} 
%    = \frac{2E}{3 (1+t_{\rm GW}/t_{\rm env})}
    = \frac{1}{3} \frac{\left(\pi f_r \right)^{\frac{2}{3}} \cM^{\frac{5}{3}}}{1+t_{\rm GW}/t_{\rm env}} \,.
\ee
When $t_{\rm GW} \ll t_{\rm env}$, we obtain the usual spectrum for GW-driven coalescence. {As the PTAs observe only a narrow range of frequencies, environmentally-driven decay can be approximated by
\be
    \frac{t_{\rm env}}{t_{\rm GW}}
    = \left(\frac{f_r}{f_{\rm GW}}\right)^{\alpha_1 + \alpha_2 \ln f_r/f_{\rm GW + \ldots} }
\ee
where $f_{\rm GW}$ denotes a reference frequency above which GW emission becomes dominant and $\alpha_n$ are coefficients series expansion of $\ln t_{\rm env}/t_{\rm GW}$ in $\ln f_r$.
%\be
%    \alpha_n \equiv \frac{1}{n!}\left.\left(\frac{\td}{\td \ln f_r}\right)^n \ln \frac{t_{\rm env}}{t_{\rm GW}}\right|_{f_r = f_{\rm GW}} \,. 
%\ee
At leading order, we can drop $\alpha_{n>1}$ terms and define $\alpha_1 \equiv \alpha$}, so that the environmental effects are parametrized by a power-law dependence on the orbital frequency, cf. \cite{Haiman:2009te, Sesana:2013wja, Kelley:2016gse, Kozhikkal:2023gkt} and choose~\footnote{This parametrization of the environmental effects does not account for the ratio between the masses of the components of the binary. However, the majority of the binaries are expected to have mass ratios close to unity~\cite{Ellis:2023owy}. {An analogous power-law model was used in the NANOGrav analysis~\cite{NANOGrav:2023gor}.} }
\bea \label{eq:gas}
    %\frac{t_{\rm env}}{t_{\rm GW}} = \left(\frac{f_r}{f_{\rm GW}}\right)^{\alpha} , \quad 
    f_{\rm GW}(\cM,\eta,z) = f_{\rm ref} \left(\frac{\cM}{10^9 \Msun}\right)^{-\beta} , 
%    t_{\rm env} = t_{\rm GW}(\cM_*,f_*) \left(\frac{\cM}{\cM_{*}}\right)^{\beta} \left(\frac{f_r}{f_{*}}\right)^{\alpha}
\eea
where $f_{\rm ref}$, $\alpha>0$ are phenomenological parameters that we constrain with the NG15 data (we note that $t_{\rm env}$ increases with $f_r$ if $\alpha>8/3$).
Depending on the environmental mechanism for orbital decay, the range $\beta = 0.2-0.8$ has been considered in the literature~\cite{Sesana:2013wja, Kozhikkal:2023gkt, Haiman:2009te}. Following Ref.~\cite{Kozhikkal:2023gkt}, we fix $\beta = 0.4$.  As the signal is generated by binaries in a relatively narrow BH mass range, we expect our results to be only weakly dependent on $\beta$.

\begin{figure}
    \centering
    \includegraphics[width=\columnwidth]{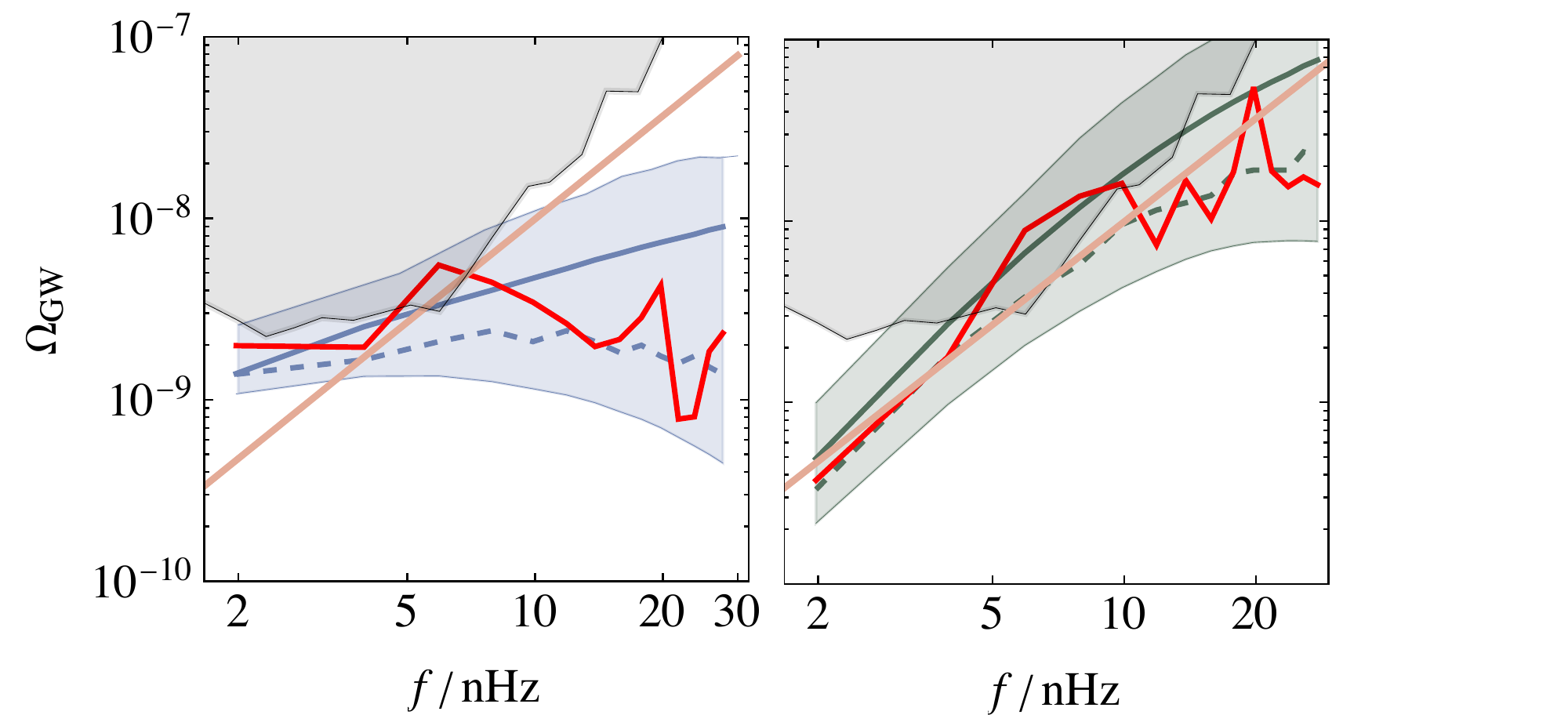}
    \caption{The best-fit GW energy density for GW-driven binaries (left) and GW and environmentally-driven effects (right), compared to the best power-law fit to the NG15 data (orange). The mean energy densities \eqref{eq:OmegaGW} are shown as full lines and the most probable values as dashed lines, and in each case one random realization is shown in red and the $95\%$ CL ranges are shaded. The gray curve shows the NG15 upper limit on signals from individually resolvable binaries~\cite{NANOGrav:2023pdq}.}
    \label{fig:column}
\end{figure}

\begin{figure}
    \centering    \includegraphics[width=0.8\columnwidth]{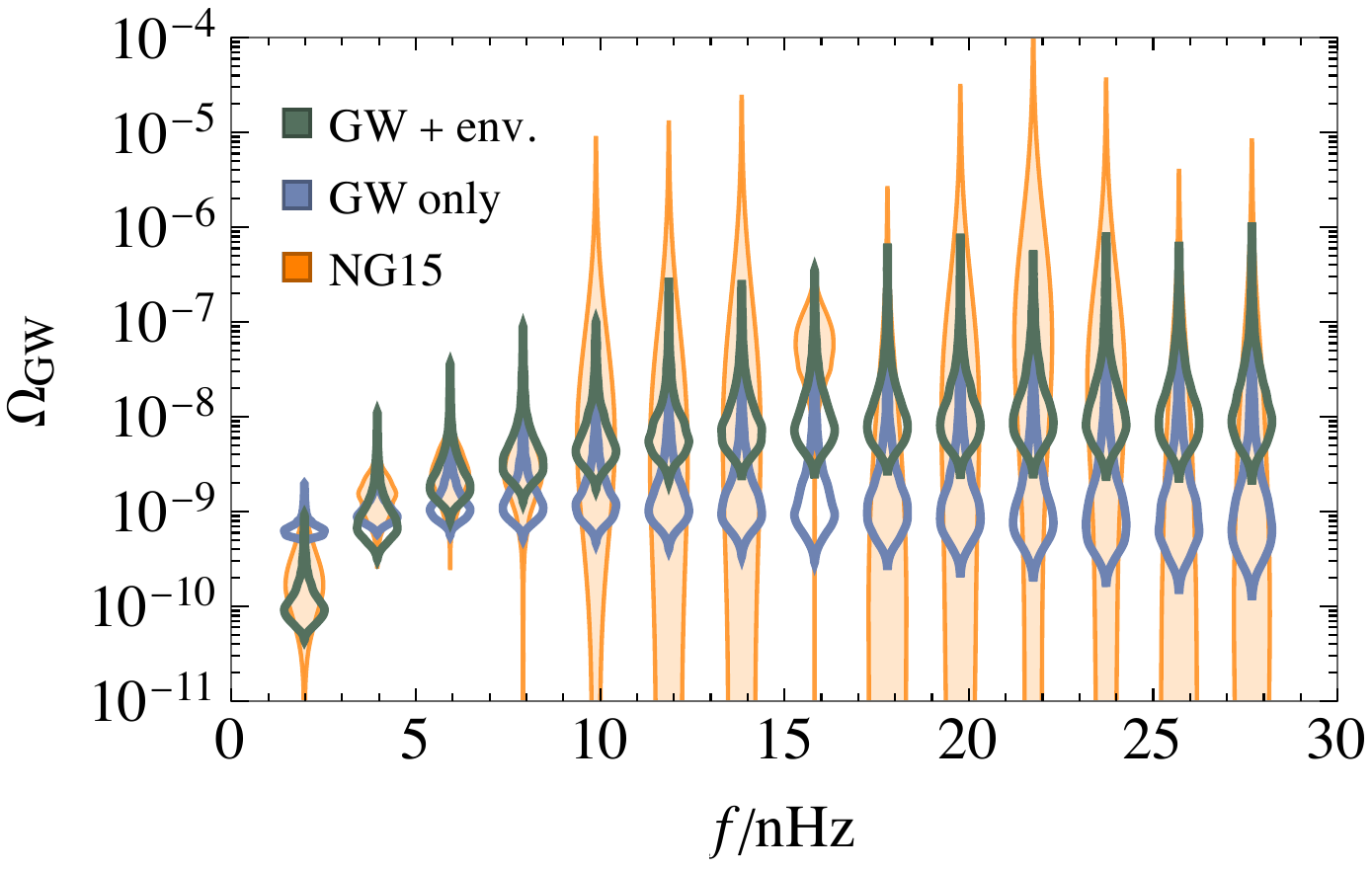}
    \caption{The NG15 data on the energy density of GWs, $\Omega_{\rm GW}$, (shown in orange), compared with the best fit assuming energy loss by GWs only (shown in blue), and the best fit including also energy loss by the environmental effects (shown in green).  The posteriors for the parameters of the fits are shown in Fig.~\ref{fig:corner}.}
    \label{fig:bins}
\end{figure}

%%%%%%%%%%%%%%%%%%%%%%%%%%%%%%%%%%%%%%%%%%%
\vspace{5pt}\noindent\textbf{Analysis --}
%%%%%%%%%%%%%%%%%%%%%%%%%%%%%%%%%%%%%%%%%%%
As reported by NANOGrav, a power-law fit of the form $\Omega_{\rm GW} = A (f/f_{\yr})^{5-\gamma}$ excludes the $\gamma = 13/3$ scaling naively expected for a background of SMBH binaries whose evolution is driven by GW emission at the $95\%$ CL.\footnote{As a consistency check, we repeated our analysis for a power-law ansatz and reproduced the posteriors reported by NANOGrav~\cite{NANOGrav:2023gor}.} However, as shown in Fig.~\ref{fig:column}, due to the stochastic nature of the signal, the power law fails to capture many of its properties and it could differ significantly from any one particular realization of the background. Moreover, due to the environmental effects, even the mean GW background is not a simple power-law. Therefore we use the full information provided in~\cite{NANOGrav:2023gor} for the probability distribution functions at each frequency bin, $P_{\rm data}(\Omega, f_i)$, {represented by the orange `violins' in Fig.~\ref{fig:bins}.}

Following~\cite{Ellis:2023owy}, we generate Monte-Carlo realizations of the stochastic GW background using the same frequency binning as the NANOGrav Collaboration, $f_i$. A realization of the SMBH background in a specific frequency bin is given by
\be\label{eq:Omega_i}
    \Omega_{\rm GW}(f_j)  = \frac{1}{\ln(f_{j+1}/f_j)}\sum_{k=1}^{N(f_j)} \Omega_{\rm GW}^{(1)}(\vec{\theta}^k_b)  \,,
\ee
where $\vec{\theta}_b \equiv \{\mathcal{M},z,\eta,f\}$ denotes the parameters of a binary and {$N(f_j)$ is drawn from a Poisson distribution determined by the expected number of binaries $\bar N(f_j) = \int_{f\in (f_{j},f_{j+1})} \!\td \lambda\, \td\tau $ contributing to each bin.} The contribution from an individual binary emitting at some frequency is
\be
    \Omega_{\rm GW}^{(1)}(f, \vec{\theta}_b) 
    = \frac{1+z}{4\pi D_L^2 \rho_c} \frac{\td E_{\rm GW}}{\td t} \, .
\ee
To decrease the computation time, rather than generating realizations of the binary population, we instead generate values of $\Omega_{\rm GW}(f_j)$ directly. Assuming that there are no correlations between the frequency bins, \ie, no binaries cross between bins during the period of observation, the statistical properties of $\Omega_{\rm GW}(f)$ can be inferred from the distribution of $\Omega_{{\rm GW}}^{(1)}(f, \vec{\theta}_b)$, which is given by\footnote{The binary inclination can be exactly accounted for by including the inclination angle dependent prefactor in $\Omega_{\rm GW}^{(1)}$ and integrating over the inclination angle in $P^{(1)}$. We have checked that including the inclination angle in $P^{(1)}$ has a negligible effect and thus we use the inclination angle averaged $\Omega_{\rm GW}^{(1)}$.}
\be\label{eq:P1}
    P^{(1)}(\Omega|f,\vec{\theta}_f) \propto \int \td \lambda \left| \frac{\td t}{\td \ln f_r} \right| 
    \delta\left( \Omega - \Omega^{(1)}_{\rm GW} \right) \, ,
  \ee
and depends only on the parameters of the merger rate and binary evolution $\vec{\theta}_f=\left(p_{\rm BH},\alpha,\beta,f_{\rm ref}\right)$. 

{Finally, the distribution $P(\Omega|f_j,\vec{\theta}_f)$ of the total GW energy density~\eqref{eq:Omega_i} in each bin is estimated by dividing the signal into two pieces: the signal from the strongest sources, and the rest (see the Supplemental Material for details). The first component is modeled using a Monte-Carlo approach, simulating $4\times10^5$ realizations of the signal from these strong sources using single event distributions~\eqref{eq:P1}. The rest of the signal follows a narrow Gaussian distribution and can be simply modeled by its average. The probability distribution of $\Omega_{\rm GW}(f_j)$ possesses a long tail which it inherits from $P^{(1)}$.~\footnote{It asymptotes to $P(\Omega|f,\vec{\theta}_f) \sim \Omega^{-\frac{5}{2}}$ as $\Omega\rightarrow \infty$ thus implying a divergent variance~\cite{Ellis:2023owy}.} Such long tails are expensive to resolve with the Monte Carlo approach, so we use $P^{(1)}$ to construct the large $\Omega$ tail analytically. This improves the statistical modeling of $\Omega_{\rm GW}(f_j)$ at large amplitudes for which a single source is expected to dominate the signal. We represent the probability distributions of $\Omega$ with and without the environmental effects for the best-fit parameter values in Fig.~\ref{fig:bins} by the green and blue `violins'.}

Our approach provides an accurate and fast~\footnote{{It takes $\mathcal{O}(1)$\ second per bin to compute $P(\Omega|f_j,\vec{\theta}_f)$ on an Apple M1 Pro 8-core processor.}} way of resolving the distribution $P(\Omega|f_j,\vec{\theta}_f)$ of spectral fluctuations in each frequency bin. This makes it feasible to perform accurate scans over a wide range of model parameters $\vec{\theta}_f$, and is thus a step forward from earlier analyses that relied on Gaussian process interpolation for SMBH population synthesis~\cite{NANOGrav:2023hfp} or a relatively small number of realizations of the full SMBH population~\cite{EPTA:2023xxk}.

The likelihood of a given model $\vec{\theta}_f$ is 
\be
    l(f_i|\vec{\theta}_f)
    \propto 
    %\prod_{i}\int \td \Omega\,  P(\Omega|f_i,\vec{\theta}_f)\, P_{\rm data}(\Omega|f_i)\, ,
    \prod_{i}\int \td \Omega\,  P_{\rm data}(f_i|\Omega) P(\Omega|f_i,\vec{\theta}_f)\, \, ,
\ee
where the effective number of variables of $\vec{\theta}_f$ depends on the hypothesis made for the population: if the energy loss due to GW emission dominates -- corresponding to very slow environmental energy loss $(t_{\rm env}\rightarrow \infty)$ -- the only parameter is $p_{\rm BH}$,
{whereas environmental energy loss is characterized by two additional parameters, $f_{\rm ref}$ and $\alpha$}. Finally, to compare the fits we define the likelihood ratio
\be
    \ell_{\rm max}=\frac{{\rm max}_{\vec{\theta}_f}l(\vec{\theta}_f |H_1)}{{\rm max}_{\vec{\theta}_f}l(\vec{\theta}_f|H_2)} \,.
\ee

%%%%%%%%%%%%%%%%%%%%%%%%%%%%%%%%%%%%%%%%%%%
\vspace{5pt}\noindent\textbf{Results --}
%%%%%%%%%%%%%%%%%%%%%%%%%%%%%%%%%%%%%%%%%%%
We see in Fig.~\ref{fig:bins} that the NG15 data (shown in orange) are better fitted by the model including environmental effects (shown in green) than by the model where binaries evolve purely by emitting GWs (shown in blue). In particular, the fit including environmental energy loss captures better the dip in the lowest frequency bin at $f = 2$\,nHz whereas the GW-driven model has to balance the tilt by lowering the overall merging efficiency. We note that suppression of the signal at low frequencies is expected also when the binaries are eccentric~\cite{Huerta:2015pva}.

\begin{figure}
    \centering
    \includegraphics[width=0.99\columnwidth]{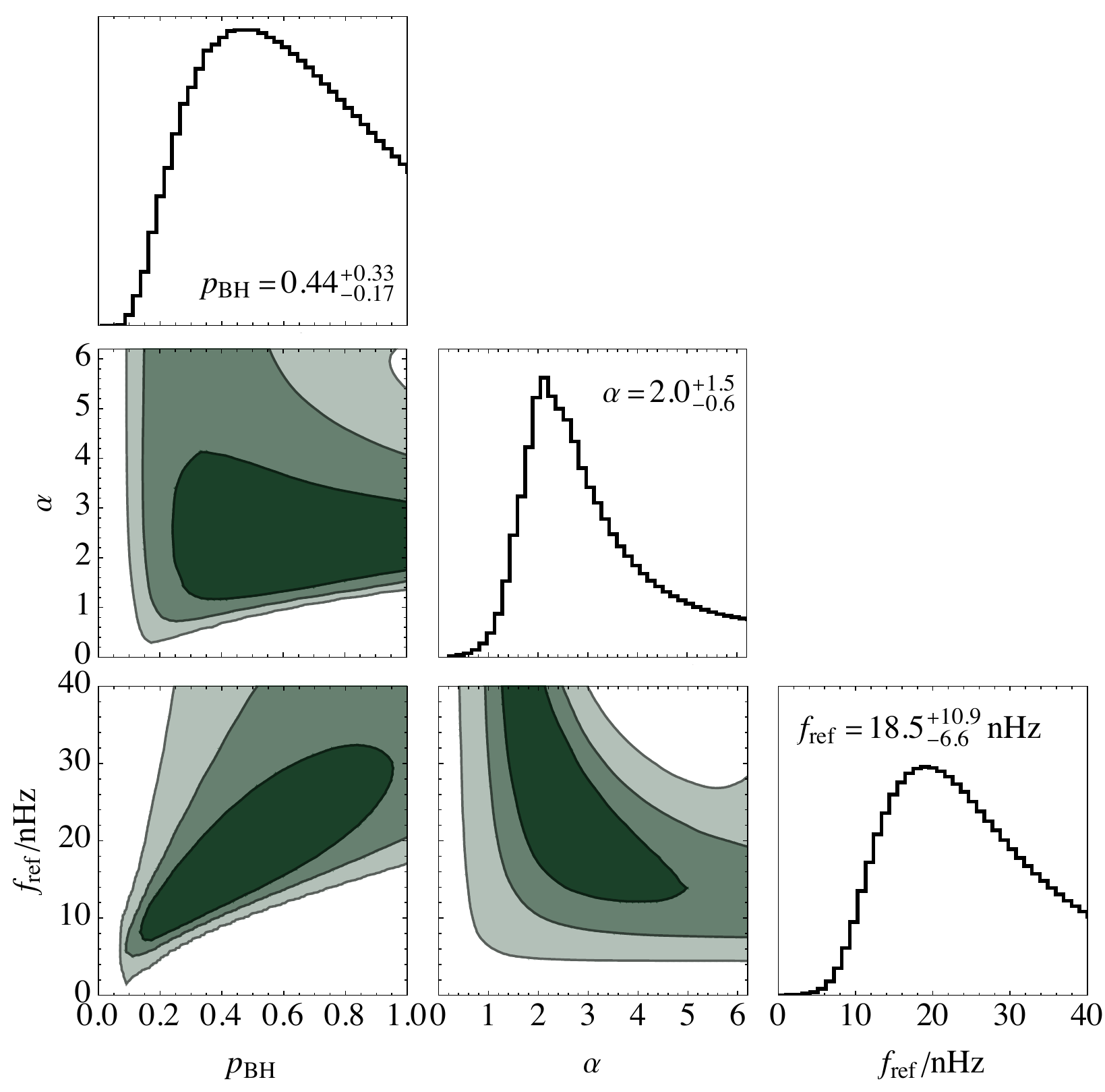}
    \caption{Fit of the SMBH binary model with environmental effects parametrized by $f_{\rm ref}$, $\alpha$, and $p_{\rm BH}$ to the NG15 data~\cite{NANOGrav:2023gor}, for $\beta=2/5$.}
    \label{fig:corner}
\end{figure}

We find that the best-fit value of the merging efficiency in the case of purely GW-driven binaries is $p_{\rm BH} = 0.07_{-0.07}^{+0.05}$, where the uncertainties are reported at the $68\%$ CL. For the model including environmental effects the fit prefers larger values of $p_{\rm BH}$. In this case, we show in Fig.~\ref{fig:corner} the two-dimensional marginalized projections of the three-dimensional parameter space featuring the 68, 95 and 99\% CL regions with different shades of green and the best fit as a white dot. The marginalized distributions of the three parameters are also shown. The best-fit values are $p_{\rm BH}=0.84$, $\alpha = 2.0$, and $f_{\rm ref} \simeq 34\, {\rm nHz}$. We note, however, that there are significant correlations between the parameters and the marginalized probability distributions are broad and asymmetric. The preferred value of $\alpha$ is consistent with previous model estimates and the value $\alpha = 8/3$ for which $t_{\rm env}$ would be frequency-independent. The central value of $f_{\rm ref}$ corresponds to the apparent break in the spectral index seen in Fig.~\ref{fig:bins}.

The model that includes environmental effects gives a larger best-fit likelihood but also has more parameters, namely $(p_{\rm BH},\alpha,f_{\rm ref})$. The likelihood ratio with respect to the best fit for the purely GW-driven binaries (i.e., the limit $f_{\rm ref}\to 0$) corresponds to $\Delta \chi^2\sim -2\log{\ell_{\rm max}} = 11$. In a Gaussian approximation, since the difference in the number of degrees of freedom between the hypotheses is 2, the model including environmental effects is favored over the model with pure GW evolution at the $2\sigma$ CL.

\begin{figure}
    \centering    
    \includegraphics[width=0.8\columnwidth]{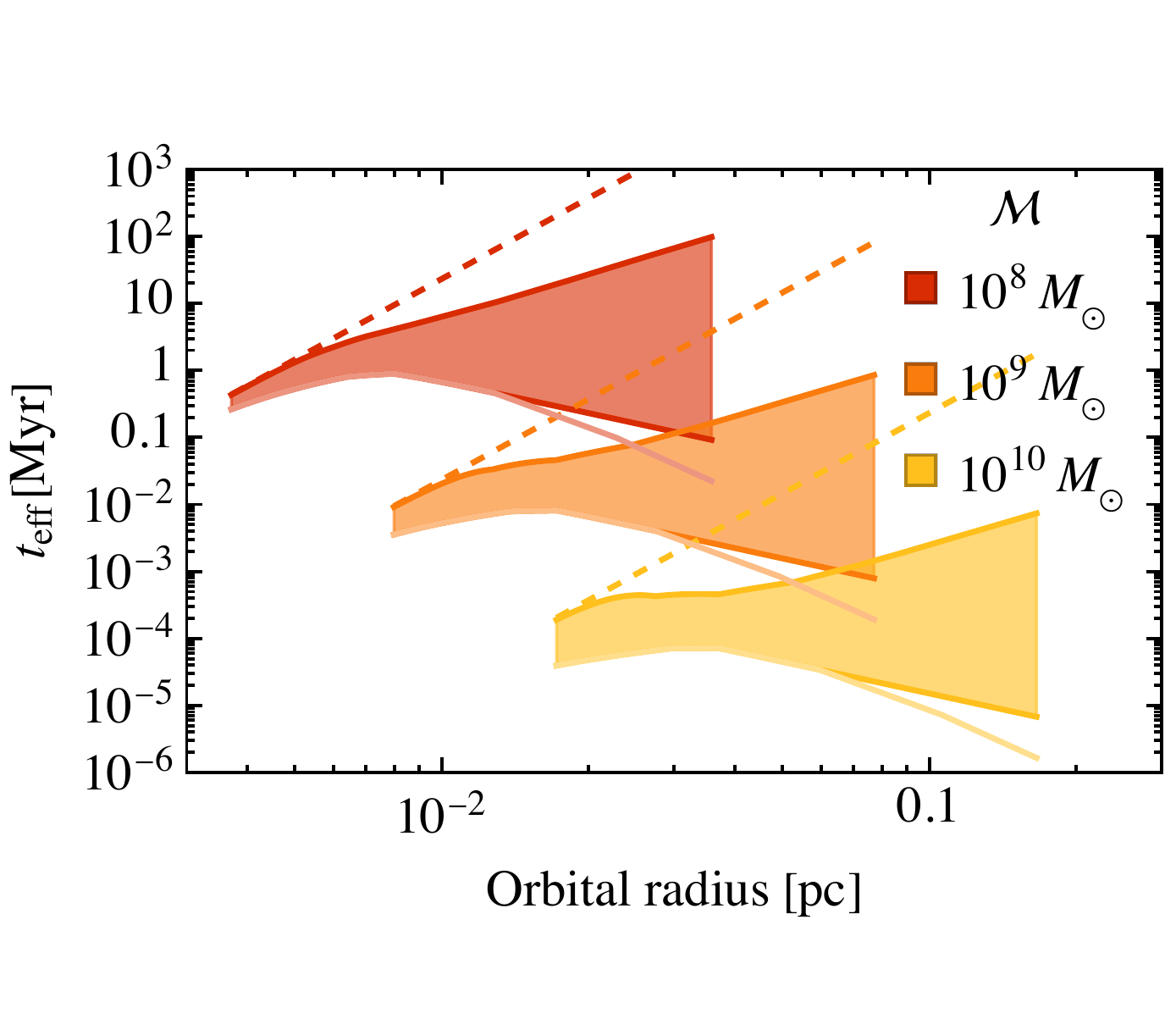}
    \vspace{-5mm}
    \caption{Effective timescales $t_{\rm eff}$ for fits to the NG15 data, shown as functions of the separations at which binaries are emitting GWs in the NANOGrav band (1\,-\,30\,nHz), and for chirp masses $\mathcal{M}\in\left(10^8, 10^{10}\right)M_{\odot}$, which is the range found in \cite{Ellis:2023owy} to dominate the stochastic GW background. The dashed lines are for energy loss by GWs alone, and the shaded regions are those favored at the 68\% CL for a combination of energy loss by environmental effects and GWs. The two lower lines are computed with $\alpha<4$ and $\alpha<6$ as priors.}
    \label{fig:effective_time}
\end{figure}

We note that the lower bound on $\Omega_{\rm GW}$ in the 8th bin at $f = 16$\,nHz is relatively high. Large upward fluctuations are a feature of the long-tailed distribution of the GW spectra from SMBHs~\cite{Ellis:2023owy}, and are not penalized heavily in a likelihood analysis. Relatively nearby SMBH binaries can generate such peaks. Specifically, the best fits for the GW-only and GW + environment models shown in Fig.~\ref{fig:column} predict that 0.44 and 0.85 SMBH binaries are expected to be resolved with the current NANOGrav sensitivity, respectively. Intriguingly, we find that in both cases the individually resolvable binaries are most likely to be found in the 4-7 nHz frequency range, which coincides with a potential monochromatic GW signal seen by NANOGrav~\cite{NANOGrav:2023pdq} and EPTA~\cite{Antoniadis:2023aac}.

%%%%%%%%%%%%%%%%%%%%%%%%%%%%%%%%%%%%%%%%%%%
\vspace{5pt}\noindent\textbf{Discussion --}
%%%%%%%%%%%%%%%%%%%%%%%%%%%%%%%%%%%%%%%%%%%
The PTA results open an exciting new way to probe the environments of SMBH binaries. We show in Fig.~\ref{fig:effective_time} the effective timescales for the evolution of the binaries that best fit the NG15 data. We see that at the smallest radii, the binaries evolve by emitting GWs, whereas at larger distances the binary timescales are close to being flat. This suggests that at these distances the main mechanism driving binary evolution is viscous drag. Simulations show that in this regime the combination with other effects, \eg, loss-cone scattering as well as GW emission leads to a distribution that is close to flat, although the uncertainties are large. In general, we find that the phenomenological fit is consistent with the expected evolution of binaries in galactic environments.

As discussed in~\cite{Ellis:2023owy}, although there are large uncertainties, extrapolation of the model predicts that GWs from mergers of lower-mass BHs should be observable in the higher frequency ranges where the LISA mission~\cite{Audley:2017drz} and proposed atom interferometers such as AION~\cite{Badurina:2019hst} and AEDGE~\cite{Bertoldi:2019tck} are most sensitive, as well as other projects targeting the deciHz frequency band~\cite{Corbin:2005ny,Nakamura:2016hna}. LISA would be most sensitive to mergers of BHs with masses $(10^3, 10^7) M_{\odot}$ and AEDGE to the $(10^2, 10^5) M_{\odot}$ mass range. Observations of binaries of such intermediate-mass BHs could cast light on the mechanism for the assembly of SMBHs as well as probe strong gravity in a range beyond the reach of terrestrial laser interferometers. To the extent that the results of this paper suggest a high merger efficiency, $p_{\rm BH} \sim 0.5$, the merger rates might be greater than was estimated in~\cite{Ellis:2023owy}, improving the prospects for seeing inspiralling intermediate-mass BH binaries by LISA and AEDGE to ${\cal O}(10^3)$ and by km-scale terrestrial atom interferometers to ${\cal O}(10)$.

\begin{acknowledgments}
\vspace{5pt}\noindent\emph{Acknowledgments --}
This work was supported by European Regional Development Fund through the CoE program grant TK133 and by the Estonian Research Council grants PSG869 and PRG803. The work of J.E. and M.F. was supported by the United Kingdom STFC Grants ST/T000759/1 and ST/T00679X/1. The work of V.V. was partially supported by the European Union's Horizon Europe research and innovation program under the Marie Sk\l{}odowska-Curie grant agreement No. 101065736.
\end{acknowledgments}

\bibliography{refs}

%merlin.mbs apsrev4-1.bst 2010-07-25 4.21a (PWD, AO, DPC) hacked
%Control: key (0)
%Control: author (8) initials jnrlst
%Control: editor formatted (1) identically to author
%Control: production of article title (-1) disabled
%Control: page (0) single
%Control: year (1) truncated
%Control: production of eprint (0) enabled
\begin{thebibliography}{100}%
\makeatletter
\providecommand \@ifxundefined [1]{%
 \@ifx{#1\undefined}
}%
\providecommand \@ifnum [1]{%
 \ifnum #1\expandafter \@firstoftwo
 \else \expandafter \@secondoftwo
 \fi
}%
\providecommand \@ifx [1]{%
 \ifx #1\expandafter \@firstoftwo
 \else \expandafter \@secondoftwo
 \fi
}%
\providecommand \natexlab [1]{#1}%
\providecommand \enquote  [1]{``#1''}%
\providecommand \bibnamefont  [1]{#1}%
\providecommand \bibfnamefont [1]{#1}%
\providecommand \citenamefont [1]{#1}%
\providecommand \href@noop [0]{\@secondoftwo}%
\providecommand \href [0]{\begingroup \@sanitize@url \@href}%
\providecommand \@href[1]{\@@startlink{#1}\@@href}%
\providecommand \@@href[1]{\endgroup#1\@@endlink}%
\providecommand \@sanitize@url [0]{\catcode `\\12\catcode `\$12\catcode
  `\&12\catcode `\#12\catcode `\^12\catcode `\_12\catcode `\%12\relax}%
\providecommand \@@startlink[1]{}%
\providecommand \@@endlink[0]{}%
\providecommand \url  [0]{\begingroup\@sanitize@url \@url }%
\providecommand \@url [1]{\endgroup\@href {#1}{\urlprefix }}%
\providecommand \urlprefix  [0]{URL }%
\providecommand \Eprint [0]{\href }%
\providecommand \doibase [0]{http://dx.doi.org/}%
\providecommand \selectlanguage [0]{\@gobble}%
\providecommand \bibinfo  [0]{\@secondoftwo}%
\providecommand \bibfield  [0]{\@secondoftwo}%
\providecommand \translation [1]{[#1]}%
\providecommand \BibitemOpen [0]{}%
\providecommand \bibitemStop [0]{}%
\providecommand \bibitemNoStop [0]{.\EOS\space}%
\providecommand \EOS [0]{\spacefactor3000\relax}%
\providecommand \BibitemShut  [1]{\csname bibitem#1\endcsname}%
\let\auto@bib@innerbib\@empty
%</preamble>
\bibitem [{\citenamefont {Agazie}\ \emph
  {et~al.}(2023{\natexlab{a}})\citenamefont {Agazie} \emph
  {et~al.}}]{NANOGrav:2023hde}%
  \BibitemOpen
  \bibfield  {author} {\bibinfo {author} {\bibfnamefont {G.}~\bibnamefont
  {Agazie}} \emph {et~al.} (\bibinfo {collaboration} {NANOGrav
  Collaboration}),\ }\href {\doibase 10.3847/2041-8213/acda9a} {\bibfield
  {journal} {\bibinfo  {journal} {Astrophys. J. Lett.}\ }\textbf {\bibinfo
  {volume} {951}},\ \bibinfo {pages} {L9} (\bibinfo {year}
  {2023}{\natexlab{a}})},\ \Eprint {http://arxiv.org/abs/2306.16217}
  {arXiv:2306.16217 [astro-ph.HE]} \BibitemShut {NoStop}%
\bibitem [{\citenamefont {Antoniadis}\ \emph
  {et~al.}(2023{\natexlab{a}})\citenamefont {Antoniadis} \emph
  {et~al.}}]{EPTA:2023sfo}%
  \BibitemOpen
  \bibfield  {author} {\bibinfo {author} {\bibfnamefont {J.}~\bibnamefont
  {Antoniadis}} \emph {et~al.} (\bibinfo {collaboration} {EPTA
  Collaboration}),\ }\href {\doibase 10.1051/0004-6361/202346841} {\  (\bibinfo
  {year} {2023}{\natexlab{a}}),\ 10.1051/0004-6361/202346841},\ \Eprint
  {http://arxiv.org/abs/2306.16224} {arXiv:2306.16224 [astro-ph.HE]}
  \BibitemShut {NoStop}%
\bibitem [{\citenamefont {Zic}\ \emph {et~al.}(2023)\citenamefont {Zic} \emph
  {et~al.}}]{Zic:2023gta}%
  \BibitemOpen
  \bibfield  {author} {\bibinfo {author} {\bibfnamefont {A.}~\bibnamefont
  {Zic}} \emph {et~al.} (\bibinfo {collaboration} {Parkes Pulsar Timing Array
  Collaboration}),\ }\href@noop {} {\  (\bibinfo {year} {2023})},\ \Eprint
  {http://arxiv.org/abs/2306.16230} {arXiv:2306.16230 [astro-ph.HE]}
  \BibitemShut {NoStop}%
\bibitem [{\citenamefont {Xu}\ \emph {et~al.}(2023)\citenamefont {Xu} \emph
  {et~al.}}]{Xu:2023wog}%
  \BibitemOpen
  \bibfield  {author} {\bibinfo {author} {\bibfnamefont {H.}~\bibnamefont {Xu}}
  \emph {et~al.} (\bibinfo {collaboration} {Chinese Pulsar Timing Array
  Collaboration}),\ }\href {\doibase 10.1088/1674-4527/acdfa5} {\  (\bibinfo
  {year} {2023}),\ 10.1088/1674-4527/acdfa5},\ \Eprint
  {http://arxiv.org/abs/2306.16216} {arXiv:2306.16216 [astro-ph.HE]}
  \BibitemShut {NoStop}%
\bibitem [{\citenamefont {Agazie}\ \emph
  {et~al.}(2023{\natexlab{b}})\citenamefont {Agazie} \emph
  {et~al.}}]{NANOGrav:2023gor}%
  \BibitemOpen
  \bibfield  {author} {\bibinfo {author} {\bibfnamefont {G.}~\bibnamefont
  {Agazie}} \emph {et~al.} (\bibinfo {collaboration} {NANOGrav
  Collaboration}),\ }\href {\doibase 10.3847/2041-8213/acdac6} {\bibfield
  {journal} {\bibinfo  {journal} {Astrophys. J. Lett.}\ }\textbf {\bibinfo
  {volume} {951}},\ \bibinfo {pages} {L8} (\bibinfo {year}
  {2023}{\natexlab{b}})},\ \Eprint {http://arxiv.org/abs/2306.16213}
  {arXiv:2306.16213 [astro-ph.HE]} \BibitemShut {NoStop}%
\bibitem [{\citenamefont {Hellings}\ and\ \citenamefont
  {Downs}(1983)}]{Hellings:1983fr}%
  \BibitemOpen
  \bibfield  {author} {\bibinfo {author} {\bibfnamefont {R.~W.}\ \bibnamefont
  {Hellings}}\ and\ \bibinfo {author} {\bibfnamefont {G.~S.}\ \bibnamefont
  {Downs}},\ }\href {\doibase 10.1086/183954} {\bibfield  {journal} {\bibinfo
  {journal} {Astrophys. J. Lett.}\ }\textbf {\bibinfo {volume} {265}},\
  \bibinfo {pages} {L39} (\bibinfo {year} {1983})}\BibitemShut {NoStop}%
\bibitem [{\citenamefont {Antoniadis}\ \emph
  {et~al.}(2023{\natexlab{b}})\citenamefont {Antoniadis} \emph
  {et~al.}}]{Antoniadis:2023ott}%
  \BibitemOpen
  \bibfield  {author} {\bibinfo {author} {\bibfnamefont {J.}~\bibnamefont
  {Antoniadis}} \emph {et~al.} (\bibinfo {collaboration} {European Pulsar
  Timing Array Collaboration}),\ }\href@noop {} {\  (\bibinfo {year}
  {2023}{\natexlab{b}})},\ \Eprint {http://arxiv.org/abs/2306.16214}
  {arXiv:2306.16214 [astro-ph.HE]} \BibitemShut {NoStop}%
\bibitem [{\citenamefont {Reardon}\ \emph {et~al.}(2023)\citenamefont {Reardon}
  \emph {et~al.}}]{Reardon:2023gzh}%
  \BibitemOpen
  \bibfield  {author} {\bibinfo {author} {\bibfnamefont {D.~J.}\ \bibnamefont
  {Reardon}} \emph {et~al.},\ }\href {\doibase 10.3847/2041-8213/acdd02}
  {\bibfield  {journal} {\bibinfo  {journal} {Astrophys. J. Lett.}\ }\textbf
  {\bibinfo {volume} {951}},\ \bibinfo {pages} {L6} (\bibinfo {year} {2023})},\
  \Eprint {http://arxiv.org/abs/2306.16215} {arXiv:2306.16215 [astro-ph.HE]}
  \BibitemShut {NoStop}%
\bibitem [{\citenamefont {Arzoumanian}\ \emph {et~al.}(2020)\citenamefont
  {Arzoumanian} \emph {et~al.}}]{Arzoumanian:2020vkk}%
  \BibitemOpen
  \bibfield  {author} {\bibinfo {author} {\bibfnamefont {Z.}~\bibnamefont
  {Arzoumanian}} \emph {et~al.} (\bibinfo {collaboration} {NANOGrav
  Collaboration}),\ }\href {\doibase 10.3847/2041-8213/abd401} {\bibfield
  {journal} {\bibinfo  {journal} {Astrophys. J. Lett.}\ }\textbf {\bibinfo
  {volume} {905}},\ \bibinfo {pages} {L34} (\bibinfo {year} {2020})},\ \Eprint
  {http://arxiv.org/abs/2009.04496} {arXiv:2009.04496 [astro-ph.HE]}
  \BibitemShut {NoStop}%
\bibitem [{\citenamefont {Goncharov}\ \emph {et~al.}(2021)\citenamefont
  {Goncharov} \emph {et~al.}}]{Goncharov:2021oub}%
  \BibitemOpen
  \bibfield  {author} {\bibinfo {author} {\bibfnamefont {B.}~\bibnamefont
  {Goncharov}} \emph {et~al.} (\bibinfo {collaboration} {Parkes Pulsar Timing
  Array Collaboration}),\ }\href {\doibase 10.3847/2041-8213/ac17f4} {\bibfield
   {journal} {\bibinfo  {journal} {Astrophys. J. Lett.}\ }\textbf {\bibinfo
  {volume} {917}},\ \bibinfo {pages} {L19} (\bibinfo {year} {2021})},\ \Eprint
  {http://arxiv.org/abs/2107.12112} {arXiv:2107.12112 [astro-ph.HE]}
  \BibitemShut {NoStop}%
\bibitem [{\citenamefont {Chen}\ \emph {et~al.}(2021)\citenamefont {Chen} \emph
  {et~al.}}]{Chen:2021rqp}%
  \BibitemOpen
  \bibfield  {author} {\bibinfo {author} {\bibfnamefont {S.}~\bibnamefont
  {Chen}} \emph {et~al.} (\bibinfo {collaboration} {European Pulsar Timing
  Array Collaboration}),\ }\href {\doibase 10.1093/mnras/stab2833} {\bibfield
  {journal} {\bibinfo  {journal} {Mon. Not. Roy. Astron. Soc.}\ }\textbf
  {\bibinfo {volume} {508}},\ \bibinfo {pages} {4970} (\bibinfo {year}
  {2021})},\ \Eprint {http://arxiv.org/abs/2110.13184} {arXiv:2110.13184
  [astro-ph.HE]} \BibitemShut {NoStop}%
\bibitem [{\citenamefont {Antoniadis}\ \emph {et~al.}(2022)\citenamefont
  {Antoniadis} \emph {et~al.}}]{Antoniadis:2022pcn}%
  \BibitemOpen
  \bibfield  {author} {\bibinfo {author} {\bibfnamefont {J.}~\bibnamefont
  {Antoniadis}} \emph {et~al.} (\bibinfo {collaboration} {International Pulsar
  Timing Array Collaboration}),\ }\href {\doibase 10.1093/mnras/stab3418}
  {\bibfield  {journal} {\bibinfo  {journal} {Mon. Not. Roy. Astron. Soc.}\
  }\textbf {\bibinfo {volume} {510}},\ \bibinfo {pages} {4873} (\bibinfo {year}
  {2022})},\ \Eprint {http://arxiv.org/abs/2201.03980} {arXiv:2201.03980
  [astro-ph.HE]} \BibitemShut {NoStop}%
\bibitem [{\citenamefont {Hulse}\ and\ \citenamefont
  {Taylor}(1975)}]{Hulse:1974eb}%
  \BibitemOpen
  \bibfield  {author} {\bibinfo {author} {\bibfnamefont {R.~A.}\ \bibnamefont
  {Hulse}}\ and\ \bibinfo {author} {\bibfnamefont {J.~H.}\ \bibnamefont
  {Taylor}},\ }\href {\doibase 10.1086/181708} {\bibfield  {journal} {\bibinfo
  {journal} {Astrophys. J. Lett.}\ }\textbf {\bibinfo {volume} {195}},\
  \bibinfo {pages} {L51} (\bibinfo {year} {1975})}\BibitemShut {NoStop}%
\bibitem [{\citenamefont {Abbott}\ \emph {et~al.}(2016)\citenamefont {Abbott}
  \emph {et~al.}}]{LIGOScientific:2016aoc}%
  \BibitemOpen
  \bibfield  {author} {\bibinfo {author} {\bibfnamefont {B.~P.}\ \bibnamefont
  {Abbott}} \emph {et~al.} (\bibinfo {collaboration} {LIGO Scientific and Virgo
  Collaborations}),\ }\href {\doibase 10.1103/PhysRevLett.116.061102}
  {\bibfield  {journal} {\bibinfo  {journal} {Phys. Rev. Lett.}\ }\textbf
  {\bibinfo {volume} {116}},\ \bibinfo {pages} {061102} (\bibinfo {year}
  {2016})},\ \Eprint {http://arxiv.org/abs/1602.03837} {arXiv:1602.03837
  [gr-qc]} \BibitemShut {NoStop}%
\bibitem [{\citenamefont {Agazie}\ \emph
  {et~al.}(2023{\natexlab{c}})\citenamefont {Agazie} \emph
  {et~al.}}]{NANOGrav:2023hfp}%
  \BibitemOpen
  \bibfield  {author} {\bibinfo {author} {\bibfnamefont {G.}~\bibnamefont
  {Agazie}} \emph {et~al.} (\bibinfo {collaboration} {NANOGrav
  Collaboration}),\ }\href {\doibase 10.3847/2041-8213/ace18b} {\bibfield
  {journal} {\bibinfo  {journal} {Astrophys. J. Lett.}\ }\textbf {\bibinfo
  {volume} {952}},\ \bibinfo {pages} {L37} (\bibinfo {year}
  {2023}{\natexlab{c}})},\ \Eprint {http://arxiv.org/abs/2306.16220}
  {arXiv:2306.16220 [astro-ph.HE]} \BibitemShut {NoStop}%
\bibitem [{\citenamefont {Antoniadis}\ \emph
  {et~al.}(2023{\natexlab{c}})\citenamefont {Antoniadis} \emph
  {et~al.}}]{Antoniadis:2023xlr}%
  \BibitemOpen
  \bibfield  {author} {\bibinfo {author} {\bibfnamefont {J.}~\bibnamefont
  {Antoniadis}} \emph {et~al.} (\bibinfo {collaboration} {European Pulsar
  Timing Array Collaboration}),\ }\href@noop {} {\  (\bibinfo {year}
  {2023}{\natexlab{c}})},\ \Eprint {http://arxiv.org/abs/2306.16227}
  {arXiv:2306.16227 [astro-ph.CO]} \BibitemShut {NoStop}%
\bibitem [{\citenamefont {Ellis}\ and\ \citenamefont
  {Lewicki}(2021)}]{Ellis:2020ena}%
  \BibitemOpen
  \bibfield  {author} {\bibinfo {author} {\bibfnamefont {J.}~\bibnamefont
  {Ellis}}\ and\ \bibinfo {author} {\bibfnamefont {M.}~\bibnamefont
  {Lewicki}},\ }\href {\doibase 10.1103/PhysRevLett.126.041304} {\bibfield
  {journal} {\bibinfo  {journal} {Phys. Rev. Lett.}\ }\textbf {\bibinfo
  {volume} {126}},\ \bibinfo {pages} {041304} (\bibinfo {year} {2021})},\
  \Eprint {http://arxiv.org/abs/2009.06555} {arXiv:2009.06555 [astro-ph.CO]}
  \BibitemShut {NoStop}%
\bibitem [{\citenamefont {Datta}\ \emph {et~al.}(2021)\citenamefont {Datta},
  \citenamefont {Ghosal},\ and\ \citenamefont {Samanta}}]{Datta:2020bht}%
  \BibitemOpen
  \bibfield  {author} {\bibinfo {author} {\bibfnamefont {S.}~\bibnamefont
  {Datta}}, \bibinfo {author} {\bibfnamefont {A.}~\bibnamefont {Ghosal}}, \
  and\ \bibinfo {author} {\bibfnamefont {R.}~\bibnamefont {Samanta}},\ }\href
  {\doibase 10.1088/1475-7516/2021/08/021} {\bibfield  {journal} {\bibinfo
  {journal} {JCAP}\ }\textbf {\bibinfo {volume} {08}},\ \bibinfo {pages} {021}
  (\bibinfo {year} {2021})},\ \Eprint {http://arxiv.org/abs/2012.14981}
  {arXiv:2012.14981 [hep-ph]} \BibitemShut {NoStop}%
\bibitem [{\citenamefont {Samanta}\ and\ \citenamefont
  {Datta}(2021)}]{Samanta:2020cdk}%
  \BibitemOpen
  \bibfield  {author} {\bibinfo {author} {\bibfnamefont {R.}~\bibnamefont
  {Samanta}}\ and\ \bibinfo {author} {\bibfnamefont {S.}~\bibnamefont
  {Datta}},\ }\href {\doibase 10.1007/JHEP05(2021)211} {\bibfield  {journal}
  {\bibinfo  {journal} {JHEP}\ }\textbf {\bibinfo {volume} {05}},\ \bibinfo
  {pages} {211} (\bibinfo {year} {2021})},\ \Eprint
  {http://arxiv.org/abs/2009.13452} {arXiv:2009.13452 [hep-ph]} \BibitemShut
  {NoStop}%
\bibitem [{\citenamefont {Buchmuller}\ \emph {et~al.}(2020)\citenamefont
  {Buchmuller}, \citenamefont {Domcke},\ and\ \citenamefont
  {Schmitz}}]{Buchmuller:2020lbh}%
  \BibitemOpen
  \bibfield  {author} {\bibinfo {author} {\bibfnamefont {W.}~\bibnamefont
  {Buchmuller}}, \bibinfo {author} {\bibfnamefont {V.}~\bibnamefont {Domcke}},
  \ and\ \bibinfo {author} {\bibfnamefont {K.}~\bibnamefont {Schmitz}},\ }\href
  {\doibase 10.1016/j.physletb.2020.135914} {\bibfield  {journal} {\bibinfo
  {journal} {Phys. Lett. B}\ }\textbf {\bibinfo {volume} {811}},\ \bibinfo
  {pages} {135914} (\bibinfo {year} {2020})},\ \Eprint
  {http://arxiv.org/abs/2009.10649} {arXiv:2009.10649 [astro-ph.CO]}
  \BibitemShut {NoStop}%
\bibitem [{\citenamefont {Blasi}\ \emph {et~al.}(2021)\citenamefont {Blasi},
  \citenamefont {Brdar},\ and\ \citenamefont {Schmitz}}]{Blasi:2020mfx}%
  \BibitemOpen
  \bibfield  {author} {\bibinfo {author} {\bibfnamefont {S.}~\bibnamefont
  {Blasi}}, \bibinfo {author} {\bibfnamefont {V.}~\bibnamefont {Brdar}}, \ and\
  \bibinfo {author} {\bibfnamefont {K.}~\bibnamefont {Schmitz}},\ }\href
  {\doibase 10.1103/PhysRevLett.126.041305} {\bibfield  {journal} {\bibinfo
  {journal} {Phys. Rev. Lett.}\ }\textbf {\bibinfo {volume} {126}},\ \bibinfo
  {pages} {041305} (\bibinfo {year} {2021})},\ \Eprint
  {http://arxiv.org/abs/2009.06607} {arXiv:2009.06607 [astro-ph.CO]}
  \BibitemShut {NoStop}%
\bibitem [{\citenamefont {Buchmuller}\ \emph {et~al.}(2021)\citenamefont
  {Buchmuller}, \citenamefont {Domcke},\ and\ \citenamefont
  {Schmitz}}]{Buchmuller:2021mbb}%
  \BibitemOpen
  \bibfield  {author} {\bibinfo {author} {\bibfnamefont {W.}~\bibnamefont
  {Buchmuller}}, \bibinfo {author} {\bibfnamefont {V.}~\bibnamefont {Domcke}},
  \ and\ \bibinfo {author} {\bibfnamefont {K.}~\bibnamefont {Schmitz}},\ }\href
  {\doibase 10.1088/1475-7516/2021/12/006} {\bibfield  {journal} {\bibinfo
  {journal} {JCAP}\ }\textbf {\bibinfo {volume} {12}},\ \bibinfo {pages} {006}
  (\bibinfo {year} {2021})},\ \Eprint {http://arxiv.org/abs/2107.04578}
  {arXiv:2107.04578 [hep-ph]} \BibitemShut {NoStop}%
\bibitem [{\citenamefont {Blanco-Pillado}\ \emph {et~al.}(2021)\citenamefont
  {Blanco-Pillado}, \citenamefont {Olum},\ and\ \citenamefont
  {Wachter}}]{Blanco-Pillado:2021ygr}%
  \BibitemOpen
  \bibfield  {author} {\bibinfo {author} {\bibfnamefont {J.~J.}\ \bibnamefont
  {Blanco-Pillado}}, \bibinfo {author} {\bibfnamefont {K.~D.}\ \bibnamefont
  {Olum}}, \ and\ \bibinfo {author} {\bibfnamefont {J.~M.}\ \bibnamefont
  {Wachter}},\ }\href {\doibase 10.1103/PhysRevD.103.103512} {\bibfield
  {journal} {\bibinfo  {journal} {Phys. Rev. D}\ }\textbf {\bibinfo {volume}
  {103}},\ \bibinfo {pages} {103512} (\bibinfo {year} {2021})},\ \Eprint
  {http://arxiv.org/abs/2102.08194} {arXiv:2102.08194 [astro-ph.CO]}
  \BibitemShut {NoStop}%
\bibitem [{\citenamefont {Ferreira}\ \emph {et~al.}(2023)\citenamefont
  {Ferreira}, \citenamefont {Notari}, \citenamefont {Pujolas},\ and\
  \citenamefont {Rompineve}}]{Ferreira:2022zzo}%
  \BibitemOpen
  \bibfield  {author} {\bibinfo {author} {\bibfnamefont {R.~Z.}\ \bibnamefont
  {Ferreira}}, \bibinfo {author} {\bibfnamefont {A.}~\bibnamefont {Notari}},
  \bibinfo {author} {\bibfnamefont {O.}~\bibnamefont {Pujolas}}, \ and\
  \bibinfo {author} {\bibfnamefont {F.}~\bibnamefont {Rompineve}},\ }\href
  {\doibase 10.1088/1475-7516/2023/02/001} {\bibfield  {journal} {\bibinfo
  {journal} {JCAP}\ }\textbf {\bibinfo {volume} {02}},\ \bibinfo {pages} {001}
  (\bibinfo {year} {2023})},\ \Eprint {http://arxiv.org/abs/2204.04228}
  {arXiv:2204.04228 [astro-ph.CO]} \BibitemShut {NoStop}%
\bibitem [{\citenamefont {An}\ and\ \citenamefont {Yang}(2023)}]{An:2023idh}%
  \BibitemOpen
  \bibfield  {author} {\bibinfo {author} {\bibfnamefont {H.}~\bibnamefont
  {An}}\ and\ \bibinfo {author} {\bibfnamefont {C.}~\bibnamefont {Yang}},\
  }\href@noop {} {\  (\bibinfo {year} {2023})},\ \Eprint
  {http://arxiv.org/abs/2304.02361} {arXiv:2304.02361 [hep-ph]} \BibitemShut
  {NoStop}%
\bibitem [{\citenamefont {Qiu}\ and\ \citenamefont {Yu}(2023)}]{Qiu:2023wbs}%
  \BibitemOpen
  \bibfield  {author} {\bibinfo {author} {\bibfnamefont {Z.-Y.}\ \bibnamefont
  {Qiu}}\ and\ \bibinfo {author} {\bibfnamefont {Z.-H.}\ \bibnamefont {Yu}},\
  }\href@noop {} {\  (\bibinfo {year} {2023})},\ \Eprint
  {http://arxiv.org/abs/2304.02506} {arXiv:2304.02506 [hep-ph]} \BibitemShut
  {NoStop}%
\bibitem [{\citenamefont {Zeng}\ \emph {et~al.}(2023)\citenamefont {Zeng},
  \citenamefont {Liu},\ and\ \citenamefont {Guo}}]{Zeng:2023jut}%
  \BibitemOpen
  \bibfield  {author} {\bibinfo {author} {\bibfnamefont {Z.-M.}\ \bibnamefont
  {Zeng}}, \bibinfo {author} {\bibfnamefont {J.}~\bibnamefont {Liu}}, \ and\
  \bibinfo {author} {\bibfnamefont {Z.-K.}\ \bibnamefont {Guo}},\ }\href@noop
  {} {\  (\bibinfo {year} {2023})},\ \Eprint {http://arxiv.org/abs/2301.07230}
  {arXiv:2301.07230 [astro-ph.CO]} \BibitemShut {NoStop}%
\bibitem [{\citenamefont {Arzoumanian}\ \emph {et~al.}(2021)\citenamefont
  {Arzoumanian} \emph {et~al.}}]{NANOGrav:2021flc}%
  \BibitemOpen
  \bibfield  {author} {\bibinfo {author} {\bibfnamefont {Z.}~\bibnamefont
  {Arzoumanian}} \emph {et~al.} (\bibinfo {collaboration} {NANOGrav
  Collaboration}),\ }\href {\doibase 10.1103/PhysRevLett.127.251302} {\bibfield
   {journal} {\bibinfo  {journal} {Phys. Rev. Lett.}\ }\textbf {\bibinfo
  {volume} {127}},\ \bibinfo {pages} {251302} (\bibinfo {year} {2021})},\
  \Eprint {http://arxiv.org/abs/2104.13930} {arXiv:2104.13930 [astro-ph.CO]}
  \BibitemShut {NoStop}%
\bibitem [{\citenamefont {Xue}\ \emph {et~al.}(2021)\citenamefont {Xue} \emph
  {et~al.}}]{Xue:2021gyq}%
  \BibitemOpen
  \bibfield  {author} {\bibinfo {author} {\bibfnamefont {X.}~\bibnamefont
  {Xue}} \emph {et~al.},\ }\href {\doibase 10.1103/PhysRevLett.127.251303}
  {\bibfield  {journal} {\bibinfo  {journal} {Phys. Rev. Lett.}\ }\textbf
  {\bibinfo {volume} {127}},\ \bibinfo {pages} {251303} (\bibinfo {year}
  {2021})},\ \Eprint {http://arxiv.org/abs/2110.03096} {arXiv:2110.03096
  [astro-ph.CO]} \BibitemShut {NoStop}%
\bibitem [{\citenamefont {Nakai}\ \emph {et~al.}(2021)\citenamefont {Nakai},
  \citenamefont {Suzuki}, \citenamefont {Takahashi},\ and\ \citenamefont
  {Yamada}}]{Nakai:2020oit}%
  \BibitemOpen
  \bibfield  {author} {\bibinfo {author} {\bibfnamefont {Y.}~\bibnamefont
  {Nakai}}, \bibinfo {author} {\bibfnamefont {M.}~\bibnamefont {Suzuki}},
  \bibinfo {author} {\bibfnamefont {F.}~\bibnamefont {Takahashi}}, \ and\
  \bibinfo {author} {\bibfnamefont {M.}~\bibnamefont {Yamada}},\ }\href
  {\doibase 10.1016/j.physletb.2021.136238} {\bibfield  {journal} {\bibinfo
  {journal} {Phys. Lett. B}\ }\textbf {\bibinfo {volume} {816}},\ \bibinfo
  {pages} {136238} (\bibinfo {year} {2021})},\ \Eprint
  {http://arxiv.org/abs/2009.09754} {arXiv:2009.09754 [astro-ph.CO]}
  \BibitemShut {NoStop}%
\bibitem [{\citenamefont {Di~Bari}\ \emph {et~al.}(2021)\citenamefont
  {Di~Bari}, \citenamefont {Marfatia},\ and\ \citenamefont
  {Zhou}}]{DiBari:2021dri}%
  \BibitemOpen
  \bibfield  {author} {\bibinfo {author} {\bibfnamefont {P.}~\bibnamefont
  {Di~Bari}}, \bibinfo {author} {\bibfnamefont {D.}~\bibnamefont {Marfatia}}, \
  and\ \bibinfo {author} {\bibfnamefont {Y.-L.}\ \bibnamefont {Zhou}},\ }\href
  {\doibase 10.1007/JHEP10(2021)193} {\bibfield  {journal} {\bibinfo  {journal}
  {JHEP}\ }\textbf {\bibinfo {volume} {10}},\ \bibinfo {pages} {193} (\bibinfo
  {year} {2021})},\ \Eprint {http://arxiv.org/abs/2106.00025} {arXiv:2106.00025
  [hep-ph]} \BibitemShut {NoStop}%
\bibitem [{\citenamefont {Sakharov}\ \emph {et~al.}(2021)\citenamefont
  {Sakharov}, \citenamefont {Eroshenko},\ and\ \citenamefont
  {Rubin}}]{Sakharov:2021dim}%
  \BibitemOpen
  \bibfield  {author} {\bibinfo {author} {\bibfnamefont {A.~S.}\ \bibnamefont
  {Sakharov}}, \bibinfo {author} {\bibfnamefont {Y.~N.}\ \bibnamefont
  {Eroshenko}}, \ and\ \bibinfo {author} {\bibfnamefont {S.~G.}\ \bibnamefont
  {Rubin}},\ }\href {\doibase 10.1103/PhysRevD.104.043005} {\bibfield
  {journal} {\bibinfo  {journal} {Phys. Rev. D}\ }\textbf {\bibinfo {volume}
  {104}},\ \bibinfo {pages} {043005} (\bibinfo {year} {2021})},\ \Eprint
  {http://arxiv.org/abs/2104.08750} {arXiv:2104.08750 [hep-ph]} \BibitemShut
  {NoStop}%
\bibitem [{\citenamefont {Li}\ \emph {et~al.}(2021)\citenamefont {Li},
  \citenamefont {Shao}, \citenamefont {Wu},\ and\ \citenamefont
  {Yu}}]{Li:2021qer}%
  \BibitemOpen
  \bibfield  {author} {\bibinfo {author} {\bibfnamefont {S.-L.}\ \bibnamefont
  {Li}}, \bibinfo {author} {\bibfnamefont {L.}~\bibnamefont {Shao}}, \bibinfo
  {author} {\bibfnamefont {P.}~\bibnamefont {Wu}}, \ and\ \bibinfo {author}
  {\bibfnamefont {H.}~\bibnamefont {Yu}},\ }\href {\doibase
  10.1103/PhysRevD.104.043510} {\bibfield  {journal} {\bibinfo  {journal}
  {Phys. Rev. D}\ }\textbf {\bibinfo {volume} {104}},\ \bibinfo {pages}
  {043510} (\bibinfo {year} {2021})},\ \Eprint
  {http://arxiv.org/abs/2101.08012} {arXiv:2101.08012 [astro-ph.CO]}
  \BibitemShut {NoStop}%
\bibitem [{\citenamefont {Ashoorioon}\ \emph {et~al.}(2022)\citenamefont
  {Ashoorioon}, \citenamefont {Rezazadeh},\ and\ \citenamefont
  {Rostami}}]{Ashoorioon:2022raz}%
  \BibitemOpen
  \bibfield  {author} {\bibinfo {author} {\bibfnamefont {A.}~\bibnamefont
  {Ashoorioon}}, \bibinfo {author} {\bibfnamefont {K.}~\bibnamefont
  {Rezazadeh}}, \ and\ \bibinfo {author} {\bibfnamefont {A.}~\bibnamefont
  {Rostami}},\ }\href {\doibase 10.1016/j.physletb.2022.137542} {\bibfield
  {journal} {\bibinfo  {journal} {Phys. Lett. B}\ }\textbf {\bibinfo {volume}
  {835}},\ \bibinfo {pages} {137542} (\bibinfo {year} {2022})},\ \Eprint
  {http://arxiv.org/abs/2202.01131} {arXiv:2202.01131 [astro-ph.CO]}
  \BibitemShut {NoStop}%
\bibitem [{\citenamefont {Benetti}\ \emph {et~al.}(2022)\citenamefont
  {Benetti}, \citenamefont {Graef},\ and\ \citenamefont
  {Vagnozzi}}]{Benetti:2021uea}%
  \BibitemOpen
  \bibfield  {author} {\bibinfo {author} {\bibfnamefont {M.}~\bibnamefont
  {Benetti}}, \bibinfo {author} {\bibfnamefont {L.~L.}\ \bibnamefont {Graef}},
  \ and\ \bibinfo {author} {\bibfnamefont {S.}~\bibnamefont {Vagnozzi}},\
  }\href {\doibase 10.1103/PhysRevD.105.043520} {\bibfield  {journal} {\bibinfo
   {journal} {Phys. Rev. D}\ }\textbf {\bibinfo {volume} {105}},\ \bibinfo
  {pages} {043520} (\bibinfo {year} {2022})},\ \Eprint
  {http://arxiv.org/abs/2111.04758} {arXiv:2111.04758 [astro-ph.CO]}
  \BibitemShut {NoStop}%
\bibitem [{\citenamefont {Barir}\ \emph {et~al.}(2022)\citenamefont {Barir},
  \citenamefont {Geller}, \citenamefont {Sun},\ and\ \citenamefont
  {Volansky}}]{Barir:2022kzo}%
  \BibitemOpen
  \bibfield  {author} {\bibinfo {author} {\bibfnamefont {J.}~\bibnamefont
  {Barir}}, \bibinfo {author} {\bibfnamefont {M.}~\bibnamefont {Geller}},
  \bibinfo {author} {\bibfnamefont {C.}~\bibnamefont {Sun}}, \ and\ \bibinfo
  {author} {\bibfnamefont {T.}~\bibnamefont {Volansky}},\ }\href@noop {} {\
  (\bibinfo {year} {2022})},\ \Eprint {http://arxiv.org/abs/2203.00693}
  {arXiv:2203.00693 [hep-ph]} \BibitemShut {NoStop}%
\bibitem [{\citenamefont {Hindmarsh}\ and\ \citenamefont
  {Kume}(2023)}]{Hindmarsh:2022awe}%
  \BibitemOpen
  \bibfield  {author} {\bibinfo {author} {\bibfnamefont {M.}~\bibnamefont
  {Hindmarsh}}\ and\ \bibinfo {author} {\bibfnamefont {J.}~\bibnamefont
  {Kume}},\ }\href {\doibase 10.1088/1475-7516/2023/04/045} {\bibfield
  {journal} {\bibinfo  {journal} {JCAP}\ }\textbf {\bibinfo {volume} {04}},\
  \bibinfo {pages} {045} (\bibinfo {year} {2023})},\ \Eprint
  {http://arxiv.org/abs/2210.06178} {arXiv:2210.06178 [astro-ph.CO]}
  \BibitemShut {NoStop}%
\bibitem [{\citenamefont {Vaskonen}\ and\ \citenamefont
  {Veerm\"ae}(2021)}]{Vaskonen:2020lbd}%
  \BibitemOpen
  \bibfield  {author} {\bibinfo {author} {\bibfnamefont {V.}~\bibnamefont
  {Vaskonen}}\ and\ \bibinfo {author} {\bibfnamefont {H.}~\bibnamefont
  {Veerm\"ae}},\ }\href {\doibase 10.1103/PhysRevLett.126.051303} {\bibfield
  {journal} {\bibinfo  {journal} {Phys. Rev. Lett.}\ }\textbf {\bibinfo
  {volume} {126}},\ \bibinfo {pages} {051303} (\bibinfo {year} {2021})},\
  \Eprint {http://arxiv.org/abs/2009.07832} {arXiv:2009.07832 [astro-ph.CO]}
  \BibitemShut {NoStop}%
\bibitem [{\citenamefont {De~Luca}\ \emph {et~al.}(2021)\citenamefont
  {De~Luca}, \citenamefont {Franciolini},\ and\ \citenamefont
  {Riotto}}]{DeLuca:2020agl}%
  \BibitemOpen
  \bibfield  {author} {\bibinfo {author} {\bibfnamefont {V.}~\bibnamefont
  {De~Luca}}, \bibinfo {author} {\bibfnamefont {G.}~\bibnamefont
  {Franciolini}}, \ and\ \bibinfo {author} {\bibfnamefont {A.}~\bibnamefont
  {Riotto}},\ }\href {\doibase 10.1103/PhysRevLett.126.041303} {\bibfield
  {journal} {\bibinfo  {journal} {Phys. Rev. Lett.}\ }\textbf {\bibinfo
  {volume} {126}},\ \bibinfo {pages} {041303} (\bibinfo {year} {2021})},\
  \Eprint {http://arxiv.org/abs/2009.08268} {arXiv:2009.08268 [astro-ph.CO]}
  \BibitemShut {NoStop}%
\bibitem [{\citenamefont {Bhaumik}\ and\ \citenamefont
  {Jain}(2021)}]{Bhaumik:2020dor}%
  \BibitemOpen
  \bibfield  {author} {\bibinfo {author} {\bibfnamefont {N.}~\bibnamefont
  {Bhaumik}}\ and\ \bibinfo {author} {\bibfnamefont {R.~K.}\ \bibnamefont
  {Jain}},\ }\href {\doibase 10.1103/PhysRevD.104.023531} {\bibfield  {journal}
  {\bibinfo  {journal} {Phys. Rev. D}\ }\textbf {\bibinfo {volume} {104}},\
  \bibinfo {pages} {023531} (\bibinfo {year} {2021})},\ \Eprint
  {http://arxiv.org/abs/2009.10424} {arXiv:2009.10424 [astro-ph.CO]}
  \BibitemShut {NoStop}%
\bibitem [{\citenamefont {Inomata}\ \emph {et~al.}(2021)\citenamefont
  {Inomata}, \citenamefont {Kawasaki}, \citenamefont {Mukaida},\ and\
  \citenamefont {Yanagida}}]{Inomata:2020xad}%
  \BibitemOpen
  \bibfield  {author} {\bibinfo {author} {\bibfnamefont {K.}~\bibnamefont
  {Inomata}}, \bibinfo {author} {\bibfnamefont {M.}~\bibnamefont {Kawasaki}},
  \bibinfo {author} {\bibfnamefont {K.}~\bibnamefont {Mukaida}}, \ and\
  \bibinfo {author} {\bibfnamefont {T.~T.}\ \bibnamefont {Yanagida}},\ }\href
  {\doibase 10.1103/PhysRevLett.126.131301} {\bibfield  {journal} {\bibinfo
  {journal} {Phys. Rev. Lett.}\ }\textbf {\bibinfo {volume} {126}},\ \bibinfo
  {pages} {131301} (\bibinfo {year} {2021})},\ \Eprint
  {http://arxiv.org/abs/2011.01270} {arXiv:2011.01270 [astro-ph.CO]}
  \BibitemShut {NoStop}%
\bibitem [{\citenamefont {Kohri}\ and\ \citenamefont
  {Terada}(2021)}]{Kohri:2020qqd}%
  \BibitemOpen
  \bibfield  {author} {\bibinfo {author} {\bibfnamefont {K.}~\bibnamefont
  {Kohri}}\ and\ \bibinfo {author} {\bibfnamefont {T.}~\bibnamefont {Terada}},\
  }\href {\doibase 10.1016/j.physletb.2020.136040} {\bibfield  {journal}
  {\bibinfo  {journal} {Phys. Lett. B}\ }\textbf {\bibinfo {volume} {813}},\
  \bibinfo {pages} {136040} (\bibinfo {year} {2021})},\ \Eprint
  {http://arxiv.org/abs/2009.11853} {arXiv:2009.11853 [astro-ph.CO]}
  \BibitemShut {NoStop}%
\bibitem [{\citenamefont {Dom\`enech}\ and\ \citenamefont
  {Pi}(2022)}]{Domenech:2020ers}%
  \BibitemOpen
  \bibfield  {author} {\bibinfo {author} {\bibfnamefont {G.}~\bibnamefont
  {Dom\`enech}}\ and\ \bibinfo {author} {\bibfnamefont {S.}~\bibnamefont
  {Pi}},\ }\href {\doibase 10.1007/s11433-021-1839-6} {\bibfield  {journal}
  {\bibinfo  {journal} {Sci. China Phys. Mech. Astron.}\ }\textbf {\bibinfo
  {volume} {65}},\ \bibinfo {pages} {230411} (\bibinfo {year} {2022})},\
  \Eprint {http://arxiv.org/abs/2010.03976} {arXiv:2010.03976 [astro-ph.CO]}
  \BibitemShut {NoStop}%
\bibitem [{\citenamefont {Vagnozzi}(2021)}]{Vagnozzi:2020gtf}%
  \BibitemOpen
  \bibfield  {author} {\bibinfo {author} {\bibfnamefont {S.}~\bibnamefont
  {Vagnozzi}},\ }\href {\doibase 10.1093/mnrasl/slaa203} {\bibfield  {journal}
  {\bibinfo  {journal} {Mon. Not. Roy. Astron. Soc.}\ }\textbf {\bibinfo
  {volume} {502}},\ \bibinfo {pages} {L11} (\bibinfo {year} {2021})},\ \Eprint
  {http://arxiv.org/abs/2009.13432} {arXiv:2009.13432 [astro-ph.CO]}
  \BibitemShut {NoStop}%
\bibitem [{\citenamefont {Namba}\ and\ \citenamefont
  {Suzuki}(2020)}]{Namba:2020kij}%
  \BibitemOpen
  \bibfield  {author} {\bibinfo {author} {\bibfnamefont {R.}~\bibnamefont
  {Namba}}\ and\ \bibinfo {author} {\bibfnamefont {M.}~\bibnamefont {Suzuki}},\
  }\href {\doibase 10.1103/PhysRevD.102.123527} {\bibfield  {journal} {\bibinfo
   {journal} {Phys. Rev. D}\ }\textbf {\bibinfo {volume} {102}},\ \bibinfo
  {pages} {123527} (\bibinfo {year} {2020})},\ \Eprint
  {http://arxiv.org/abs/2009.13909} {arXiv:2009.13909 [astro-ph.CO]}
  \BibitemShut {NoStop}%
\bibitem [{\citenamefont {Sugiyama}\ \emph {et~al.}(2021)\citenamefont
  {Sugiyama}, \citenamefont {Takhistov}, \citenamefont {Vitagliano},
  \citenamefont {Kusenko}, \citenamefont {Sasaki},\ and\ \citenamefont
  {Takada}}]{Sugiyama:2020roc}%
  \BibitemOpen
  \bibfield  {author} {\bibinfo {author} {\bibfnamefont {S.}~\bibnamefont
  {Sugiyama}}, \bibinfo {author} {\bibfnamefont {V.}~\bibnamefont {Takhistov}},
  \bibinfo {author} {\bibfnamefont {E.}~\bibnamefont {Vitagliano}}, \bibinfo
  {author} {\bibfnamefont {A.}~\bibnamefont {Kusenko}}, \bibinfo {author}
  {\bibfnamefont {M.}~\bibnamefont {Sasaki}}, \ and\ \bibinfo {author}
  {\bibfnamefont {M.}~\bibnamefont {Takada}},\ }\href {\doibase
  10.1016/j.physletb.2021.136097} {\bibfield  {journal} {\bibinfo  {journal}
  {Phys. Lett. B}\ }\textbf {\bibinfo {volume} {814}},\ \bibinfo {pages}
  {136097} (\bibinfo {year} {2021})},\ \Eprint
  {http://arxiv.org/abs/2010.02189} {arXiv:2010.02189 [astro-ph.CO]}
  \BibitemShut {NoStop}%
\bibitem [{\citenamefont {Zhou}\ \emph {et~al.}(2020)\citenamefont {Zhou},
  \citenamefont {Jiang}, \citenamefont {Cai}, \citenamefont {Sasaki},\ and\
  \citenamefont {Pi}}]{Zhou:2020kkf}%
  \BibitemOpen
  \bibfield  {author} {\bibinfo {author} {\bibfnamefont {Z.}~\bibnamefont
  {Zhou}}, \bibinfo {author} {\bibfnamefont {J.}~\bibnamefont {Jiang}},
  \bibinfo {author} {\bibfnamefont {Y.-F.}\ \bibnamefont {Cai}}, \bibinfo
  {author} {\bibfnamefont {M.}~\bibnamefont {Sasaki}}, \ and\ \bibinfo {author}
  {\bibfnamefont {S.}~\bibnamefont {Pi}},\ }\href {\doibase
  10.1103/PhysRevD.102.103527} {\bibfield  {journal} {\bibinfo  {journal}
  {Phys. Rev. D}\ }\textbf {\bibinfo {volume} {102}},\ \bibinfo {pages}
  {103527} (\bibinfo {year} {2020})},\ \Eprint
  {http://arxiv.org/abs/2010.03537} {arXiv:2010.03537 [astro-ph.CO]}
  \BibitemShut {NoStop}%
\bibitem [{\citenamefont {Lin}\ \emph {et~al.}(2023)\citenamefont {Lin},
  \citenamefont {Gao}, \citenamefont {Gong}, \citenamefont {Lu}, \citenamefont
  {Wang},\ and\ \citenamefont {Zhang}}]{Lin:2021vwc}%
  \BibitemOpen
  \bibfield  {author} {\bibinfo {author} {\bibfnamefont {J.}~\bibnamefont
  {Lin}}, \bibinfo {author} {\bibfnamefont {S.}~\bibnamefont {Gao}}, \bibinfo
  {author} {\bibfnamefont {Y.}~\bibnamefont {Gong}}, \bibinfo {author}
  {\bibfnamefont {Y.}~\bibnamefont {Lu}}, \bibinfo {author} {\bibfnamefont
  {Z.}~\bibnamefont {Wang}}, \ and\ \bibinfo {author} {\bibfnamefont
  {F.}~\bibnamefont {Zhang}},\ }\href {\doibase 10.1103/PhysRevD.107.043517}
  {\bibfield  {journal} {\bibinfo  {journal} {Phys. Rev. D}\ }\textbf {\bibinfo
  {volume} {107}},\ \bibinfo {pages} {043517} (\bibinfo {year} {2023})},\
  \Eprint {http://arxiv.org/abs/2111.01362} {arXiv:2111.01362 [gr-qc]}
  \BibitemShut {NoStop}%
\bibitem [{\citenamefont {Rezazadeh}\ \emph {et~al.}(2022)\citenamefont
  {Rezazadeh}, \citenamefont {Teimoori}, \citenamefont {Karimi},\ and\
  \citenamefont {Karami}}]{Rezazadeh:2021clf}%
  \BibitemOpen
  \bibfield  {author} {\bibinfo {author} {\bibfnamefont {K.}~\bibnamefont
  {Rezazadeh}}, \bibinfo {author} {\bibfnamefont {Z.}~\bibnamefont {Teimoori}},
  \bibinfo {author} {\bibfnamefont {S.}~\bibnamefont {Karimi}}, \ and\ \bibinfo
  {author} {\bibfnamefont {K.}~\bibnamefont {Karami}},\ }\href {\doibase
  10.1140/epjc/s10052-022-10735-w} {\bibfield  {journal} {\bibinfo  {journal}
  {Eur. Phys. J. C}\ }\textbf {\bibinfo {volume} {82}},\ \bibinfo {pages} {758}
  (\bibinfo {year} {2022})},\ \Eprint {http://arxiv.org/abs/2110.01482}
  {arXiv:2110.01482 [gr-qc]} \BibitemShut {NoStop}%
\bibitem [{\citenamefont {Kawasaki}\ and\ \citenamefont
  {Nakatsuka}(2021)}]{Kawasaki:2021ycf}%
  \BibitemOpen
  \bibfield  {author} {\bibinfo {author} {\bibfnamefont {M.}~\bibnamefont
  {Kawasaki}}\ and\ \bibinfo {author} {\bibfnamefont {H.}~\bibnamefont
  {Nakatsuka}},\ }\href {\doibase 10.1088/1475-7516/2021/05/023} {\bibfield
  {journal} {\bibinfo  {journal} {JCAP}\ }\textbf {\bibinfo {volume} {05}},\
  \bibinfo {pages} {023} (\bibinfo {year} {2021})},\ \Eprint
  {http://arxiv.org/abs/2101.11244} {arXiv:2101.11244 [astro-ph.CO]}
  \BibitemShut {NoStop}%
\bibitem [{\citenamefont {Ahmed}\ \emph {et~al.}(2022)\citenamefont {Ahmed},
  \citenamefont {Junaid},\ and\ \citenamefont {Zubair}}]{Ahmed:2021ucx}%
  \BibitemOpen
  \bibfield  {author} {\bibinfo {author} {\bibfnamefont {W.}~\bibnamefont
  {Ahmed}}, \bibinfo {author} {\bibfnamefont {M.}~\bibnamefont {Junaid}}, \
  and\ \bibinfo {author} {\bibfnamefont {U.}~\bibnamefont {Zubair}},\ }\href
  {\doibase 10.1016/j.nuclphysb.2022.115968} {\bibfield  {journal} {\bibinfo
  {journal} {Nucl. Phys. B}\ }\textbf {\bibinfo {volume} {984}},\ \bibinfo
  {pages} {115968} (\bibinfo {year} {2022})},\ \Eprint
  {http://arxiv.org/abs/2109.14838} {arXiv:2109.14838 [astro-ph.CO]}
  \BibitemShut {NoStop}%
\bibitem [{\citenamefont {Yi}\ and\ \citenamefont {Fei}(2023)}]{Yi:2022ymw}%
  \BibitemOpen
  \bibfield  {author} {\bibinfo {author} {\bibfnamefont {Z.}~\bibnamefont
  {Yi}}\ and\ \bibinfo {author} {\bibfnamefont {Q.}~\bibnamefont {Fei}},\
  }\href {\doibase 10.1140/epjc/s10052-023-11233-3} {\bibfield  {journal}
  {\bibinfo  {journal} {Eur. Phys. J. C}\ }\textbf {\bibinfo {volume} {83}},\
  \bibinfo {pages} {82} (\bibinfo {year} {2023})},\ \Eprint
  {http://arxiv.org/abs/2210.03641} {arXiv:2210.03641 [astro-ph.CO]}
  \BibitemShut {NoStop}%
\bibitem [{\citenamefont {Yi}(2023)}]{Yi:2022anu}%
  \BibitemOpen
  \bibfield  {author} {\bibinfo {author} {\bibfnamefont {Z.}~\bibnamefont
  {Yi}},\ }\href {\doibase 10.1088/1475-7516/2023/03/048} {\bibfield  {journal}
  {\bibinfo  {journal} {JCAP}\ }\textbf {\bibinfo {volume} {03}},\ \bibinfo
  {pages} {048} (\bibinfo {year} {2023})},\ \Eprint
  {http://arxiv.org/abs/2206.01039} {arXiv:2206.01039 [gr-qc]} \BibitemShut
  {NoStop}%
\bibitem [{\citenamefont {Dandoy}\ \emph {et~al.}(2023)\citenamefont {Dandoy},
  \citenamefont {Domcke},\ and\ \citenamefont {Rompineve}}]{Dandoy:2023jot}%
  \BibitemOpen
  \bibfield  {author} {\bibinfo {author} {\bibfnamefont {V.}~\bibnamefont
  {Dandoy}}, \bibinfo {author} {\bibfnamefont {V.}~\bibnamefont {Domcke}}, \
  and\ \bibinfo {author} {\bibfnamefont {F.}~\bibnamefont {Rompineve}},\
  }\href@noop {} {\  (\bibinfo {year} {2023})},\ \Eprint
  {http://arxiv.org/abs/2302.07901} {arXiv:2302.07901 [astro-ph.CO]}
  \BibitemShut {NoStop}%
\bibitem [{\citenamefont {Zhao}\ \emph {et~al.}(2023)\citenamefont {Zhao},
  \citenamefont {Liu},\ and\ \citenamefont {Li}}]{Zhao:2023xnh}%
  \BibitemOpen
  \bibfield  {author} {\bibinfo {author} {\bibfnamefont {J.-X.}\ \bibnamefont
  {Zhao}}, \bibinfo {author} {\bibfnamefont {X.-H.}\ \bibnamefont {Liu}}, \
  and\ \bibinfo {author} {\bibfnamefont {N.}~\bibnamefont {Li}},\ }\href
  {\doibase 10.1103/PhysRevD.107.043515} {\bibfield  {journal} {\bibinfo
  {journal} {Phys. Rev. D}\ }\textbf {\bibinfo {volume} {107}},\ \bibinfo
  {pages} {043515} (\bibinfo {year} {2023})},\ \Eprint
  {http://arxiv.org/abs/2302.06886} {arXiv:2302.06886 [astro-ph.CO]}
  \BibitemShut {NoStop}%
\bibitem [{\citenamefont {Ferrante}\ \emph {et~al.}(2023)\citenamefont
  {Ferrante}, \citenamefont {Franciolini}, \citenamefont {Iovino~Junior},\ and\
  \citenamefont {Urbano}}]{Ferrante:2023bgz}%
  \BibitemOpen
  \bibfield  {author} {\bibinfo {author} {\bibfnamefont {G.}~\bibnamefont
  {Ferrante}}, \bibinfo {author} {\bibfnamefont {G.}~\bibnamefont
  {Franciolini}}, \bibinfo {author} {\bibfnamefont {A.}~\bibnamefont
  {Iovino~Junior}}, \ and\ \bibinfo {author} {\bibfnamefont {A.}~\bibnamefont
  {Urbano}},\ }\href@noop {} {\  (\bibinfo {year} {2023})},\ \Eprint
  {http://arxiv.org/abs/2305.13382} {arXiv:2305.13382 [astro-ph.CO]}
  \BibitemShut {NoStop}%
\bibitem [{\citenamefont {Cai}\ \emph {et~al.}(2023)\citenamefont {Cai},
  \citenamefont {Zhu},\ and\ \citenamefont {Piao}}]{Cai:2023uhc}%
  \BibitemOpen
  \bibfield  {author} {\bibinfo {author} {\bibfnamefont {Y.}~\bibnamefont
  {Cai}}, \bibinfo {author} {\bibfnamefont {M.}~\bibnamefont {Zhu}}, \ and\
  \bibinfo {author} {\bibfnamefont {Y.-S.}\ \bibnamefont {Piao}},\ }\href@noop
  {} {\  (\bibinfo {year} {2023})},\ \Eprint {http://arxiv.org/abs/2305.10933}
  {arXiv:2305.10933 [gr-qc]} \BibitemShut {NoStop}%
\bibitem [{\citenamefont {Afzal}\ \emph {et~al.}(2023)\citenamefont {Afzal}
  \emph {et~al.}}]{NANOGrav:2023hvm}%
  \BibitemOpen
  \bibfield  {author} {\bibinfo {author} {\bibfnamefont {A.}~\bibnamefont
  {Afzal}} \emph {et~al.} (\bibinfo {collaboration} {NANOGrav Collaboration}),\
  }\href {\doibase 10.3847/2041-8213/acdc91} {\bibfield  {journal} {\bibinfo
  {journal} {Astrophys. J. Lett.}\ }\textbf {\bibinfo {volume} {951}},\
  \bibinfo {pages} {L11} (\bibinfo {year} {2023})},\ \Eprint
  {http://arxiv.org/abs/2306.16219} {arXiv:2306.16219 [astro-ph.HE]}
  \BibitemShut {NoStop}%
\bibitem [{\citenamefont {Mayer}(2013)}]{Mayer:2013jja}%
  \BibitemOpen
  \bibfield  {author} {\bibinfo {author} {\bibfnamefont {L.}~\bibnamefont
  {Mayer}},\ }\href {\doibase 10.1088/0264-9381/30/24/244008} {\bibfield
  {journal} {\bibinfo  {journal} {Class. Quant. Grav.}\ }\textbf {\bibinfo
  {volume} {30}},\ \bibinfo {pages} {244008} (\bibinfo {year} {2013})},\
  \Eprint {http://arxiv.org/abs/1308.0431} {arXiv:1308.0431 [astro-ph.CO]}
  \BibitemShut {NoStop}%
\bibitem [{\citenamefont {Foreman}\ \emph {et~al.}(2009)\citenamefont
  {Foreman}, \citenamefont {Volonteri},\ and\ \citenamefont
  {Dotti}}]{Foreman:2008th}%
  \BibitemOpen
  \bibfield  {author} {\bibinfo {author} {\bibfnamefont {G.}~\bibnamefont
  {Foreman}}, \bibinfo {author} {\bibfnamefont {M.}~\bibnamefont {Volonteri}},
  \ and\ \bibinfo {author} {\bibfnamefont {M.}~\bibnamefont {Dotti}},\ }\href
  {\doibase 10.1088/0004-637X/693/2/1554} {\bibfield  {journal} {\bibinfo
  {journal} {Astrophys. J.}\ }\textbf {\bibinfo {volume} {693}},\ \bibinfo
  {pages} {1554} (\bibinfo {year} {2009})},\ \Eprint
  {http://arxiv.org/abs/0812.1569} {arXiv:0812.1569 [astro-ph]} \BibitemShut
  {NoStop}%
\bibitem [{\citenamefont {Begelman}\ \emph {et~al.}(1980)\citenamefont
  {Begelman}, \citenamefont {Blandford},\ and\ \citenamefont
  {Rees}}]{Begelman:1980vb}%
  \BibitemOpen
  \bibfield  {author} {\bibinfo {author} {\bibfnamefont {M.~C.}\ \bibnamefont
  {Begelman}}, \bibinfo {author} {\bibfnamefont {R.~D.}\ \bibnamefont
  {Blandford}}, \ and\ \bibinfo {author} {\bibfnamefont {M.~J.}\ \bibnamefont
  {Rees}},\ }\href {\doibase 10.1038/287307a0} {\bibfield  {journal} {\bibinfo
  {journal} {Nature}\ }\textbf {\bibinfo {volume} {287}},\ \bibinfo {pages}
  {307} (\bibinfo {year} {1980})}\BibitemShut {NoStop}%
\bibitem [{\citenamefont {Phinney}(2001)}]{Phinney:2001di}%
  \BibitemOpen
  \bibfield  {author} {\bibinfo {author} {\bibfnamefont {E.~S.}\ \bibnamefont
  {Phinney}},\ }\href@noop {} {\  (\bibinfo {year} {2001})},\ \Eprint
  {http://arxiv.org/abs/astro-ph/0108028} {arXiv:astro-ph/0108028} \BibitemShut
  {NoStop}%
\bibitem [{\citenamefont {Ellis}\ \emph {et~al.}(2023)\citenamefont {Ellis},
  \citenamefont {Fairbairn}, \citenamefont {H\"utsi}, \citenamefont {Raidal},
  \citenamefont {Urrutia}, \citenamefont {Vaskonen},\ and\ \citenamefont
  {Veerm\"ae}}]{Ellis:2023owy}%
  \BibitemOpen
  \bibfield  {author} {\bibinfo {author} {\bibfnamefont {J.}~\bibnamefont
  {Ellis}}, \bibinfo {author} {\bibfnamefont {M.}~\bibnamefont {Fairbairn}},
  \bibinfo {author} {\bibfnamefont {G.}~\bibnamefont {H\"utsi}}, \bibinfo
  {author} {\bibfnamefont {M.}~\bibnamefont {Raidal}}, \bibinfo {author}
  {\bibfnamefont {J.}~\bibnamefont {Urrutia}}, \bibinfo {author} {\bibfnamefont
  {V.}~\bibnamefont {Vaskonen}}, \ and\ \bibinfo {author} {\bibfnamefont
  {H.}~\bibnamefont {Veerm\"ae}},\ }\href {\doibase
  10.1051/0004-6361/202346268} {\bibfield  {journal} {\bibinfo  {journal}
  {Astron. Astrophys.}\ }\textbf {\bibinfo {volume} {676}},\ \bibinfo {pages}
  {A38} (\bibinfo {year} {2023})},\ \Eprint {http://arxiv.org/abs/2301.13854}
  {arXiv:2301.13854 [astro-ph.CO]} \BibitemShut {NoStop}%
\bibitem [{\citenamefont {Press}\ and\ \citenamefont
  {Schechter}(1974)}]{Press:1973iz}%
  \BibitemOpen
  \bibfield  {author} {\bibinfo {author} {\bibfnamefont {W.~H.}\ \bibnamefont
  {Press}}\ and\ \bibinfo {author} {\bibfnamefont {P.}~\bibnamefont
  {Schechter}},\ }\href {\doibase 10.1086/152650} {\bibfield  {journal}
  {\bibinfo  {journal} {Astrophys. J.}\ }\textbf {\bibinfo {volume} {187}},\
  \bibinfo {pages} {425} (\bibinfo {year} {1974})}\BibitemShut {NoStop}%
\bibitem [{\citenamefont {Bond}\ \emph {et~al.}(1991)\citenamefont {Bond},
  \citenamefont {Cole}, \citenamefont {Efstathiou},\ and\ \citenamefont
  {Kaiser}}]{Bond:1990iw}%
  \BibitemOpen
  \bibfield  {author} {\bibinfo {author} {\bibfnamefont {J.~R.}\ \bibnamefont
  {Bond}}, \bibinfo {author} {\bibfnamefont {S.}~\bibnamefont {Cole}}, \bibinfo
  {author} {\bibfnamefont {G.}~\bibnamefont {Efstathiou}}, \ and\ \bibinfo
  {author} {\bibfnamefont {N.}~\bibnamefont {Kaiser}},\ }\href {\doibase
  10.1086/170520} {\bibfield  {journal} {\bibinfo  {journal} {Astrophys. J.}\
  }\textbf {\bibinfo {volume} {379}},\ \bibinfo {pages} {440} (\bibinfo {year}
  {1991})}\BibitemShut {NoStop}%
\bibitem [{\citenamefont {Lacey}\ and\ \citenamefont
  {Cole}(1993)}]{Lacey:1993iv}%
  \BibitemOpen
  \bibfield  {author} {\bibinfo {author} {\bibfnamefont {C.~G.}\ \bibnamefont
  {Lacey}}\ and\ \bibinfo {author} {\bibfnamefont {S.}~\bibnamefont {Cole}},\
  }\href@noop {} {\bibfield  {journal} {\bibinfo  {journal} {Mon. Not. Roy.
  Astron. Soc.}\ }\textbf {\bibinfo {volume} {262}},\ \bibinfo {pages} {627}
  (\bibinfo {year} {1993})}\BibitemShut {NoStop}%
\bibitem [{\citenamefont {Kormendy}\ and\ \citenamefont
  {Ho}(2013)}]{Kormendy:2013dxa}%
  \BibitemOpen
  \bibfield  {author} {\bibinfo {author} {\bibfnamefont {J.}~\bibnamefont
  {Kormendy}}\ and\ \bibinfo {author} {\bibfnamefont {L.~C.}\ \bibnamefont
  {Ho}},\ }\href {\doibase 10.1146/annurev-astro-082708-101811} {\bibfield
  {journal} {\bibinfo  {journal} {Ann. Rev. Astron. Astrophys.}\ }\textbf
  {\bibinfo {volume} {51}},\ \bibinfo {pages} {511} (\bibinfo {year} {2013})},\
  \Eprint {http://arxiv.org/abs/1304.7762} {arXiv:1304.7762 [astro-ph.CO]}
  \BibitemShut {NoStop}%
\bibitem [{\citenamefont {{Reines}}\ and\ \citenamefont
  {{Volonteri}}(2015)}]{2015ApJ...813...82R}%
  \BibitemOpen
  \bibfield  {author} {\bibinfo {author} {\bibfnamefont {A.~E.}\ \bibnamefont
  {{Reines}}}\ and\ \bibinfo {author} {\bibfnamefont {M.}~\bibnamefont
  {{Volonteri}}},\ }\href {\doibase 10.1088/0004-637X/813/2/82} {\bibfield
  {journal} {\bibinfo  {journal} {\apj}\ }\textbf {\bibinfo {volume} {813}},\
  \bibinfo {eid} {82} (\bibinfo {year} {2015})},\ \Eprint
  {http://arxiv.org/abs/1508.06274} {arXiv:1508.06274 [astro-ph.GA]}
  \BibitemShut {NoStop}%
\bibitem [{\citenamefont {Girelli}\ \emph {et~al.}(2020)\citenamefont
  {Girelli}, \citenamefont {Pozzetti}, \citenamefont {Bolzonella},
  \citenamefont {Giocoli}, \citenamefont {Marulli},\ and\ \citenamefont
  {Baldi}}]{Girelli:2020goz}%
  \BibitemOpen
  \bibfield  {author} {\bibinfo {author} {\bibfnamefont {G.}~\bibnamefont
  {Girelli}}, \bibinfo {author} {\bibfnamefont {L.}~\bibnamefont {Pozzetti}},
  \bibinfo {author} {\bibfnamefont {M.}~\bibnamefont {Bolzonella}}, \bibinfo
  {author} {\bibfnamefont {C.}~\bibnamefont {Giocoli}}, \bibinfo {author}
  {\bibfnamefont {F.}~\bibnamefont {Marulli}}, \ and\ \bibinfo {author}
  {\bibfnamefont {M.}~\bibnamefont {Baldi}},\ }\href {\doibase
  10.1051/0004-6361/201936329} {\bibfield  {journal} {\bibinfo  {journal}
  {Astron. Astrophys.}\ }\textbf {\bibinfo {volume} {634}},\ \bibinfo {pages}
  {A135} (\bibinfo {year} {2020})},\ \Eprint {http://arxiv.org/abs/2001.02230}
  {arXiv:2001.02230 [astro-ph.CO]} \BibitemShut {NoStop}%
\bibitem [{\citenamefont {Antoniadis}\ \emph
  {et~al.}(2023{\natexlab{d}})\citenamefont {Antoniadis} \emph
  {et~al.}}]{EPTA:2023xxk}%
  \BibitemOpen
  \bibfield  {author} {\bibinfo {author} {\bibfnamefont {J.}~\bibnamefont
  {Antoniadis}} \emph {et~al.} (\bibinfo {collaboration} {EPTA}),\ }\href@noop
  {} {\  (\bibinfo {year} {2023}{\natexlab{d}})},\ \Eprint
  {http://arxiv.org/abs/2306.16227} {arXiv:2306.16227 [astro-ph.CO]}
  \BibitemShut {NoStop}%
\bibitem [{\citenamefont {Enoki}\ and\ \citenamefont
  {Nagashima}(2007)}]{Enoki:2006kj}%
  \BibitemOpen
  \bibfield  {author} {\bibinfo {author} {\bibfnamefont {M.}~\bibnamefont
  {Enoki}}\ and\ \bibinfo {author} {\bibfnamefont {M.}~\bibnamefont
  {Nagashima}},\ }\href {\doibase 10.1143/PTP.117.241} {\bibfield  {journal}
  {\bibinfo  {journal} {Prog. Theor. Phys.}\ }\textbf {\bibinfo {volume}
  {117}},\ \bibinfo {pages} {241} (\bibinfo {year} {2007})},\ \Eprint
  {http://arxiv.org/abs/astro-ph/0609377} {arXiv:astro-ph/0609377} \BibitemShut
  {NoStop}%
\bibitem [{\citenamefont {Kelley}\ \emph
  {et~al.}(2017{\natexlab{a}})\citenamefont {Kelley}, \citenamefont {Blecha},
  \citenamefont {Hernquist}, \citenamefont {Sesana},\ and\ \citenamefont
  {Taylor}}]{Kelley:2017lek}%
  \BibitemOpen
  \bibfield  {author} {\bibinfo {author} {\bibfnamefont {L.~Z.}\ \bibnamefont
  {Kelley}}, \bibinfo {author} {\bibfnamefont {L.}~\bibnamefont {Blecha}},
  \bibinfo {author} {\bibfnamefont {L.}~\bibnamefont {Hernquist}}, \bibinfo
  {author} {\bibfnamefont {A.}~\bibnamefont {Sesana}}, \ and\ \bibinfo {author}
  {\bibfnamefont {S.~R.}\ \bibnamefont {Taylor}},\ }\href {\doibase
  10.1093/mnras/stx1638} {\bibfield  {journal} {\bibinfo  {journal} {Mon. Not.
  Roy. Astron. Soc.}\ }\textbf {\bibinfo {volume} {471}},\ \bibinfo {pages}
  {4508} (\bibinfo {year} {2017}{\natexlab{a}})},\ \Eprint
  {http://arxiv.org/abs/1702.02180} {arXiv:1702.02180 [astro-ph.HE]}
  \BibitemShut {NoStop}%
\bibitem [{\citenamefont {Merritt}(2013)}]{Merritt:2013awa}%
  \BibitemOpen
  \bibfield  {author} {\bibinfo {author} {\bibfnamefont {D.}~\bibnamefont
  {Merritt}},\ }\href {\doibase 10.1088/0264-9381/30/24/244005} {\bibfield
  {journal} {\bibinfo  {journal} {Class. Quant. Grav.}\ }\textbf {\bibinfo
  {volume} {30}},\ \bibinfo {pages} {244005} (\bibinfo {year} {2013})},\
  \Eprint {http://arxiv.org/abs/1307.3268} {arXiv:1307.3268 [astro-ph.GA]}
  \BibitemShut {NoStop}%
\bibitem [{\citenamefont {Armitage}\ and\ \citenamefont
  {Natarajan}(2002)}]{Armitage:2002uu}%
  \BibitemOpen
  \bibfield  {author} {\bibinfo {author} {\bibfnamefont {P.~J.}\ \bibnamefont
  {Armitage}}\ and\ \bibinfo {author} {\bibfnamefont {P.}~\bibnamefont
  {Natarajan}},\ }\href {\doibase 10.1086/339770} {\bibfield  {journal}
  {\bibinfo  {journal} {Astrophys. J. Lett.}\ }\textbf {\bibinfo {volume}
  {567}},\ \bibinfo {pages} {L9} (\bibinfo {year} {2002})},\ \Eprint
  {http://arxiv.org/abs/astro-ph/0201318} {arXiv:astro-ph/0201318} \BibitemShut
  {NoStop}%
\bibitem [{\citenamefont {Macfadyen}\ and\ \citenamefont
  {Milosavljevic}(2008)}]{Macfadyen:2006jx}%
  \BibitemOpen
  \bibfield  {author} {\bibinfo {author} {\bibfnamefont {A.~I.}\ \bibnamefont
  {Macfadyen}}\ and\ \bibinfo {author} {\bibfnamefont {M.}~\bibnamefont
  {Milosavljevic}},\ }\href {\doibase 10.1086/523869} {\bibfield  {journal}
  {\bibinfo  {journal} {Astrophys. J.}\ }\textbf {\bibinfo {volume} {672}},\
  \bibinfo {pages} {83} (\bibinfo {year} {2008})},\ \Eprint
  {http://arxiv.org/abs/astro-ph/0607467} {arXiv:astro-ph/0607467} \BibitemShut
  {NoStop}%
\bibitem [{\citenamefont {Tang}\ \emph {et~al.}(2017)\citenamefont {Tang},
  \citenamefont {MacFadyen},\ and\ \citenamefont {Haiman}}]{Tang:2017eiz}%
  \BibitemOpen
  \bibfield  {author} {\bibinfo {author} {\bibfnamefont {Y.}~\bibnamefont
  {Tang}}, \bibinfo {author} {\bibfnamefont {A.}~\bibnamefont {MacFadyen}}, \
  and\ \bibinfo {author} {\bibfnamefont {Z.}~\bibnamefont {Haiman}},\ }\href
  {\doibase 10.1093/mnras/stx1130} {\bibfield  {journal} {\bibinfo  {journal}
  {Mon. Not. Roy. Astron. Soc.}\ }\textbf {\bibinfo {volume} {469}},\ \bibinfo
  {pages} {4258} (\bibinfo {year} {2017})},\ \Eprint
  {http://arxiv.org/abs/1703.03913} {arXiv:1703.03913 [astro-ph.HE]}
  \BibitemShut {NoStop}%
\bibitem [{\citenamefont {Mu\~noz}\ \emph {et~al.}(2019)\citenamefont
  {Mu\~noz}, \citenamefont {Miranda},\ and\ \citenamefont
  {Lai}}]{Munoz:2018tnj}%
  \BibitemOpen
  \bibfield  {author} {\bibinfo {author} {\bibfnamefont {D.~J.}\ \bibnamefont
  {Mu\~noz}}, \bibinfo {author} {\bibfnamefont {R.}~\bibnamefont {Miranda}}, \
  and\ \bibinfo {author} {\bibfnamefont {D.}~\bibnamefont {Lai}},\ }\href
  {\doibase 10.3847/1538-4357/aaf867} {\bibfield  {journal} {\bibinfo
  {journal} {Astrophys. J.}\ }\textbf {\bibinfo {volume} {871}},\ \bibinfo
  {pages} {84} (\bibinfo {year} {2019})},\ \Eprint
  {http://arxiv.org/abs/1810.04676} {arXiv:1810.04676 [astro-ph.HE]}
  \BibitemShut {NoStop}%
\bibitem [{\citenamefont {Moody}\ \emph {et~al.}(2019)\citenamefont {Moody},
  \citenamefont {Shi},\ and\ \citenamefont {Stone}}]{Moody:2019nes}%
  \BibitemOpen
  \bibfield  {author} {\bibinfo {author} {\bibfnamefont {M.~S.~L.}\
  \bibnamefont {Moody}}, \bibinfo {author} {\bibfnamefont {J.-M.}\ \bibnamefont
  {Shi}}, \ and\ \bibinfo {author} {\bibfnamefont {J.~M.}\ \bibnamefont
  {Stone}},\ }\href {\doibase 10.3847/1538-4357/ab09ee} {\bibfield  {journal}
  {\bibinfo  {journal} {Astrophys. J.}\ }\textbf {\bibinfo {volume} {875}},\
  \bibinfo {pages} {66} (\bibinfo {year} {2019})},\ \Eprint
  {http://arxiv.org/abs/1903.00008} {arXiv:1903.00008 [astro-ph.HE]}
  \BibitemShut {NoStop}%
\bibitem [{\citenamefont {Tiede}\ \emph {et~al.}(2020)\citenamefont {Tiede},
  \citenamefont {Zrake}, \citenamefont {MacFadyen},\ and\ \citenamefont
  {Haiman}}]{Tiede:2020ldm}%
  \BibitemOpen
  \bibfield  {author} {\bibinfo {author} {\bibfnamefont {C.}~\bibnamefont
  {Tiede}}, \bibinfo {author} {\bibfnamefont {J.}~\bibnamefont {Zrake}},
  \bibinfo {author} {\bibfnamefont {A.}~\bibnamefont {MacFadyen}}, \ and\
  \bibinfo {author} {\bibfnamefont {Z.}~\bibnamefont {Haiman}},\ }\href
  {\doibase 10.3847/1538-4357/aba432} {\bibfield  {journal} {\bibinfo
  {journal} {Astrophys. J.}\ }\textbf {\bibinfo {volume} {900}},\ \bibinfo
  {pages} {43} (\bibinfo {year} {2020})},\ \Eprint
  {http://arxiv.org/abs/2005.09555} {arXiv:2005.09555 [astro-ph.GA]}
  \BibitemShut {NoStop}%
\bibitem [{\citenamefont {D'Orazio}\ and\ \citenamefont
  {Duffell}(2021)}]{DOrazio:2021kob}%
  \BibitemOpen
  \bibfield  {author} {\bibinfo {author} {\bibfnamefont {D.~J.}\ \bibnamefont
  {D'Orazio}}\ and\ \bibinfo {author} {\bibfnamefont {P.~C.}\ \bibnamefont
  {Duffell}},\ }\href {\doibase 10.3847/2041-8213/ac0621} {\bibfield  {journal}
  {\bibinfo  {journal} {Astrophys. J. Lett.}\ }\textbf {\bibinfo {volume}
  {914}},\ \bibinfo {pages} {L21} (\bibinfo {year} {2021})},\ \Eprint
  {http://arxiv.org/abs/2103.09251} {arXiv:2103.09251 [astro-ph.HE]}
  \BibitemShut {NoStop}%
\bibitem [{\citenamefont {Duffell}\ \emph {et~al.}(2020)\citenamefont
  {Duffell}, \citenamefont {D'Orazio}, \citenamefont {Derdzinski},
  \citenamefont {Haiman}, \citenamefont {MacFadyen}, \citenamefont {Rosen},\
  and\ \citenamefont {Zrake}}]{Duffell:2019uuk}%
  \BibitemOpen
  \bibfield  {author} {\bibinfo {author} {\bibfnamefont {P.~C.}\ \bibnamefont
  {Duffell}}, \bibinfo {author} {\bibfnamefont {D.}~\bibnamefont {D'Orazio}},
  \bibinfo {author} {\bibfnamefont {A.}~\bibnamefont {Derdzinski}}, \bibinfo
  {author} {\bibfnamefont {Z.}~\bibnamefont {Haiman}}, \bibinfo {author}
  {\bibfnamefont {A.}~\bibnamefont {MacFadyen}}, \bibinfo {author}
  {\bibfnamefont {A.~L.}\ \bibnamefont {Rosen}}, \ and\ \bibinfo {author}
  {\bibfnamefont {J.}~\bibnamefont {Zrake}},\ }\href {\doibase
  10.3847/1538-4357/abab95} {\bibfield  {journal} {\bibinfo  {journal}
  {Astrophys. J.}\ }\textbf {\bibinfo {volume} {901}},\ \bibinfo {pages} {25}
  (\bibinfo {year} {2020})},\ \Eprint {http://arxiv.org/abs/1911.05506}
  {arXiv:1911.05506 [astro-ph.SR]} \BibitemShut {NoStop}%
\bibitem [{\citenamefont {Haiman}\ \emph {et~al.}(2009)\citenamefont {Haiman},
  \citenamefont {Haiman}, \citenamefont {Kocsis}, \citenamefont {Kocsis},
  \citenamefont {Menou},\ and\ \citenamefont {Menou}}]{Haiman:2009te}%
  \BibitemOpen
  \bibfield  {author} {\bibinfo {author} {\bibfnamefont {Z.}~\bibnamefont
  {Haiman}}, \bibinfo {author} {\bibfnamefont {Z.}~\bibnamefont {Haiman}},
  \bibinfo {author} {\bibfnamefont {B.}~\bibnamefont {Kocsis}}, \bibinfo
  {author} {\bibfnamefont {B.}~\bibnamefont {Kocsis}}, \bibinfo {author}
  {\bibfnamefont {K.}~\bibnamefont {Menou}}, \ and\ \bibinfo {author}
  {\bibfnamefont {K.}~\bibnamefont {Menou}},\ }\href {\doibase
  10.1088/0004-637X/700/2/1952} {\bibfield  {journal} {\bibinfo  {journal}
  {Astrophys. J.}\ }\textbf {\bibinfo {volume} {700}},\ \bibinfo {pages} {1952}
  (\bibinfo {year} {2009})},\ \bibinfo {note} {[Erratum: Astrophys.J. 937, 129
  (2022)]},\ \Eprint {http://arxiv.org/abs/0904.1383} {arXiv:0904.1383
  [astro-ph.CO]} \BibitemShut {NoStop}%
\bibitem [{\citenamefont {Sesana}(2013)}]{Sesana:2013wja}%
  \BibitemOpen
  \bibfield  {author} {\bibinfo {author} {\bibfnamefont {A.}~\bibnamefont
  {Sesana}},\ }\href {\doibase 10.1088/0264-9381/30/22/224014} {\bibfield
  {journal} {\bibinfo  {journal} {Class. Quant. Grav.}\ }\textbf {\bibinfo
  {volume} {30}},\ \bibinfo {pages} {224014} (\bibinfo {year} {2013})},\
  \Eprint {http://arxiv.org/abs/1307.2600} {arXiv:1307.2600 [astro-ph.CO]}
  \BibitemShut {NoStop}%
\bibitem [{\citenamefont {Kelley}\ \emph
  {et~al.}(2017{\natexlab{b}})\citenamefont {Kelley}, \citenamefont {Blecha},\
  and\ \citenamefont {Hernquist}}]{Kelley:2016gse}%
  \BibitemOpen
  \bibfield  {author} {\bibinfo {author} {\bibfnamefont {L.~Z.}\ \bibnamefont
  {Kelley}}, \bibinfo {author} {\bibfnamefont {L.}~\bibnamefont {Blecha}}, \
  and\ \bibinfo {author} {\bibfnamefont {L.}~\bibnamefont {Hernquist}},\ }\href
  {\doibase 10.1093/mnras/stw2452} {\bibfield  {journal} {\bibinfo  {journal}
  {Mon. Not. Roy. Astron. Soc.}\ }\textbf {\bibinfo {volume} {464}},\ \bibinfo
  {pages} {3131} (\bibinfo {year} {2017}{\natexlab{b}})},\ \Eprint
  {http://arxiv.org/abs/1606.01900} {arXiv:1606.01900 [astro-ph.HE]}
  \BibitemShut {NoStop}%
\bibitem [{\citenamefont {Kozhikkal}\ \emph {et~al.}(2023)\citenamefont
  {Kozhikkal}, \citenamefont {Chen}, \citenamefont {Theureau}, \citenamefont
  {Habouzit},\ and\ \citenamefont {Sesana}}]{Kozhikkal:2023gkt}%
  \BibitemOpen
  \bibfield  {author} {\bibinfo {author} {\bibfnamefont {M.~M.}\ \bibnamefont
  {Kozhikkal}}, \bibinfo {author} {\bibfnamefont {S.}~\bibnamefont {Chen}},
  \bibinfo {author} {\bibfnamefont {G.}~\bibnamefont {Theureau}}, \bibinfo
  {author} {\bibfnamefont {M.}~\bibnamefont {Habouzit}}, \ and\ \bibinfo
  {author} {\bibfnamefont {A.}~\bibnamefont {Sesana}},\ }\href@noop {} {\
  (\bibinfo {year} {2023})},\ \Eprint {http://arxiv.org/abs/2305.18293}
  {arXiv:2305.18293 [astro-ph.CO]} \BibitemShut {NoStop}%
\bibitem [{\citenamefont {Agazie}\ \emph
  {et~al.}(2023{\natexlab{d}})\citenamefont {Agazie} \emph
  {et~al.}}]{NANOGrav:2023pdq}%
  \BibitemOpen
  \bibfield  {author} {\bibinfo {author} {\bibfnamefont {G.}~\bibnamefont
  {Agazie}} \emph {et~al.} (\bibinfo {collaboration} {NANOGrav
  Collaboration}),\ }\href {\doibase 10.3847/2041-8213/ace18a} {\bibfield
  {journal} {\bibinfo  {journal} {Astrophys. J. Lett.}\ }\textbf {\bibinfo
  {volume} {951}},\ \bibinfo {pages} {L50} (\bibinfo {year}
  {2023}{\natexlab{d}})},\ \Eprint {http://arxiv.org/abs/2306.16222}
  {arXiv:2306.16222 [astro-ph.HE]} \BibitemShut {NoStop}%
\bibitem [{\citenamefont {Huerta}\ \emph {et~al.}(2015)\citenamefont {Huerta},
  \citenamefont {McWilliams}, \citenamefont {Gair},\ and\ \citenamefont
  {Taylor}}]{Huerta:2015pva}%
  \BibitemOpen
  \bibfield  {author} {\bibinfo {author} {\bibfnamefont {E.~A.}\ \bibnamefont
  {Huerta}}, \bibinfo {author} {\bibfnamefont {S.~T.}\ \bibnamefont
  {McWilliams}}, \bibinfo {author} {\bibfnamefont {J.~R.}\ \bibnamefont
  {Gair}}, \ and\ \bibinfo {author} {\bibfnamefont {S.~R.}\ \bibnamefont
  {Taylor}},\ }\href {\doibase 10.1103/PhysRevD.92.063010} {\bibfield
  {journal} {\bibinfo  {journal} {Phys. Rev. D}\ }\textbf {\bibinfo {volume}
  {92}},\ \bibinfo {pages} {063010} (\bibinfo {year} {2015})},\ \Eprint
  {http://arxiv.org/abs/1504.00928} {arXiv:1504.00928 [gr-qc]} \BibitemShut
  {NoStop}%
\bibitem [{\citenamefont {Antoniadis}\ \emph
  {et~al.}(2023{\natexlab{e}})\citenamefont {Antoniadis} \emph
  {et~al.}}]{Antoniadis:2023aac}%
  \BibitemOpen
  \bibfield  {author} {\bibinfo {author} {\bibfnamefont {J.}~\bibnamefont
  {Antoniadis}} \emph {et~al.} (\bibinfo {collaboration} {EPTA
  Collaboration}),\ }\href@noop {} {\  (\bibinfo {year}
  {2023}{\natexlab{e}})},\ \Eprint {http://arxiv.org/abs/2306.16226}
  {arXiv:2306.16226 [astro-ph.HE]} \BibitemShut {NoStop}%
\bibitem [{\citenamefont {Amaro-Seoane}\ \emph {et~al.}(2017)\citenamefont
  {Amaro-Seoane} \emph {et~al.}}]{Audley:2017drz}%
  \BibitemOpen
  \bibfield  {author} {\bibinfo {author} {\bibfnamefont {P.}~\bibnamefont
  {Amaro-Seoane}} \emph {et~al.} (\bibinfo {collaboration}
  {LISA~Collaboration}),\ }\href@noop {} {\  (\bibinfo {year} {2017})},\
  \Eprint {http://arxiv.org/abs/1702.00786} {arXiv:1702.00786 [astro-ph.IM]}
  \BibitemShut {NoStop}%
\bibitem [{\citenamefont {Badurina}\ \emph {et~al.}(2020)\citenamefont
  {Badurina} \emph {et~al.}}]{Badurina:2019hst}%
  \BibitemOpen
  \bibfield  {author} {\bibinfo {author} {\bibfnamefont {L.}~\bibnamefont
  {Badurina}} \emph {et~al.} (\bibinfo {collaboration} {AION Collaboration}),\
  }\href {\doibase 10.1088/1475-7516/2020/05/011} {\bibfield  {journal}
  {\bibinfo  {journal} {JCAP}\ }\textbf {\bibinfo {volume} {05}},\ \bibinfo
  {pages} {011} (\bibinfo {year} {2020})},\ \Eprint
  {http://arxiv.org/abs/1911.11755} {arXiv:1911.11755 [astro-ph.CO]}
  \BibitemShut {NoStop}%
\bibitem [{\citenamefont {El-Neaj}\ \emph {et~al.}(2020)\citenamefont {El-Neaj}
  \emph {et~al.}}]{Bertoldi:2019tck}%
  \BibitemOpen
  \bibfield  {author} {\bibinfo {author} {\bibfnamefont {Y.~A.}\ \bibnamefont
  {El-Neaj}} \emph {et~al.} (\bibinfo {collaboration} {AEDGE Collaboration}),\
  }\href {\doibase 10.1140/epjqt/s40507-020-0080-0} {\bibfield  {journal}
  {\bibinfo  {journal} {EPJ Quant. Technol.}\ }\textbf {\bibinfo {volume}
  {7}},\ \bibinfo {pages} {6} (\bibinfo {year} {2020})},\ \Eprint
  {http://arxiv.org/abs/1908.00802} {arXiv:1908.00802 [gr-qc]} \BibitemShut
  {NoStop}%
\bibitem [{\citenamefont {Corbin}\ and\ \citenamefont
  {Cornish}(2006)}]{Corbin:2005ny}%
  \BibitemOpen
  \bibfield  {author} {\bibinfo {author} {\bibfnamefont {V.}~\bibnamefont
  {Corbin}}\ and\ \bibinfo {author} {\bibfnamefont {N.~J.}\ \bibnamefont
  {Cornish}},\ }\href {\doibase 10.1088/0264-9381/23/7/014} {\bibfield
  {journal} {\bibinfo  {journal} {Class. Quant. Grav.}\ }\textbf {\bibinfo
  {volume} {23}},\ \bibinfo {pages} {2435} (\bibinfo {year} {2006})},\ \Eprint
  {http://arxiv.org/abs/gr-qc/0512039} {arXiv:gr-qc/0512039} \BibitemShut
  {NoStop}%
\bibitem [{\citenamefont {Nakamura}\ \emph {et~al.}(2016)\citenamefont
  {Nakamura} \emph {et~al.}}]{Nakamura:2016hna}%
  \BibitemOpen
  \bibfield  {author} {\bibinfo {author} {\bibfnamefont {T.}~\bibnamefont
  {Nakamura}} \emph {et~al.},\ }\href {\doibase 10.1093/ptep/ptw127} {\bibfield
   {journal} {\bibinfo  {journal} {PTEP}\ }\textbf {\bibinfo {volume} {2016}},\
  \bibinfo {pages} {093E01} (\bibinfo {year} {2016})},\ \Eprint
  {http://arxiv.org/abs/1607.00897} {arXiv:1607.00897 [astro-ph.HE]}
  \BibitemShut {NoStop}%
\bibitem [{\citenamefont {{Harikane}}\ \emph {et~al.}(2023)\citenamefont
  {{Harikane}} \emph {et~al.}}]{2023arXiv230311946H}%
  \BibitemOpen
  \bibfield  {author} {\bibinfo {author} {\bibfnamefont {Y.}~\bibnamefont
  {{Harikane}}} \emph {et~al.},\ }\href {\doibase 10.48550/arXiv.2303.11946}
  {\bibfield  {journal} {\bibinfo  {journal} {arXiv e-prints}\ ,\ \bibinfo
  {eid} {arXiv:2303.11946}} (\bibinfo {year} {2023})},\ \Eprint
  {http://arxiv.org/abs/2303.11946} {arXiv:2303.11946 [astro-ph.GA]}
  \BibitemShut {NoStop}%
\bibitem [{\citenamefont {Kocevski}\ \emph {et~al.}(2023)\citenamefont
  {Kocevski} \emph {et~al.}}]{2023ApJ...954L...4K}%
  \BibitemOpen
  \bibfield  {author} {\bibinfo {author} {\bibfnamefont {D.~D.}\ \bibnamefont
  {Kocevski}} \emph {et~al.},\ }\href {\doibase 10.3847/2041-8213/ace5a0}
  {\bibfield  {journal} {\bibinfo  {journal} {\apjl}\ }\textbf {\bibinfo
  {volume} {954}},\ \bibinfo {eid} {L4} (\bibinfo {year} {2023})},\ \Eprint
  {http://arxiv.org/abs/2302.00012} {arXiv:2302.00012 [astro-ph.GA]}
  \BibitemShut {NoStop}%
\bibitem [{\citenamefont {Maiolino}\ \emph {et~al.}(2023)\citenamefont
  {Maiolino} \emph {et~al.}}]{Maiolino:2023bpi}%
  \BibitemOpen
  \bibfield  {author} {\bibinfo {author} {\bibfnamefont {R.}~\bibnamefont
  {Maiolino}} \emph {et~al.},\ }\href@noop {} {\  (\bibinfo {year} {2023})},\
  \Eprint {http://arxiv.org/abs/2308.01230} {arXiv:2308.01230 [astro-ph.GA]}
  \BibitemShut {NoStop}%
\bibitem [{\citenamefont {Pacucci}\ \emph {et~al.}(2023)\citenamefont
  {Pacucci}, \citenamefont {Nguyen}, \citenamefont {Carniani}, \citenamefont
  {Maiolino},\ and\ \citenamefont {Fan}}]{Pacucci:2023oci}%
  \BibitemOpen
  \bibfield  {author} {\bibinfo {author} {\bibfnamefont {F.}~\bibnamefont
  {Pacucci}}, \bibinfo {author} {\bibfnamefont {B.}~\bibnamefont {Nguyen}},
  \bibinfo {author} {\bibfnamefont {S.}~\bibnamefont {Carniani}}, \bibinfo
  {author} {\bibfnamefont {R.}~\bibnamefont {Maiolino}}, \ and\ \bibinfo
  {author} {\bibfnamefont {X.}~\bibnamefont {Fan}},\ }\href@noop {} {\
  (\bibinfo {year} {2023})},\ \Eprint {http://arxiv.org/abs/2308.12331}
  {arXiv:2308.12331 [astro-ph.GA]} \BibitemShut {NoStop}%
\bibitem [{\citenamefont {Aversa}\ \emph {et~al.}(2015)\citenamefont {Aversa},
  \citenamefont {Lapi}, \citenamefont {de~Zotti}, \citenamefont {Shankar},\
  and\ \citenamefont {Danese}}]{Aversa:2015bya}%
  \BibitemOpen
  \bibfield  {author} {\bibinfo {author} {\bibfnamefont {R.}~\bibnamefont
  {Aversa}}, \bibinfo {author} {\bibfnamefont {A.}~\bibnamefont {Lapi}},
  \bibinfo {author} {\bibfnamefont {G.}~\bibnamefont {de~Zotti}}, \bibinfo
  {author} {\bibfnamefont {F.}~\bibnamefont {Shankar}}, \ and\ \bibinfo
  {author} {\bibfnamefont {L.}~\bibnamefont {Danese}},\ }\href {\doibase
  10.1088/0004-637X/810/1/74} {\bibfield  {journal} {\bibinfo  {journal}
  {Astrophys. J.}\ }\textbf {\bibinfo {volume} {810}},\ \bibinfo {pages} {74}
  (\bibinfo {year} {2015})},\ \Eprint {http://arxiv.org/abs/1507.07318}
  {arXiv:1507.07318 [astro-ph.GA]} \BibitemShut {NoStop}%
\bibitem [{\citenamefont {Delvecchio}\ and\ \citenamefont
  {other}(2020)}]{2020ApJ...892...17D}%
  \BibitemOpen
  \bibfield  {author} {\bibinfo {author} {\bibfnamefont {I.}~\bibnamefont
  {Delvecchio}}\ and\ \bibinfo {author} {\bibnamefont {other}},\ }\href
  {\doibase 10.3847/1538-4357/ab789c} {\bibfield  {journal} {\bibinfo
  {journal} {\apj}\ }\textbf {\bibinfo {volume} {892}},\ \bibinfo {eid} {17}
  (\bibinfo {year} {2020})},\ \Eprint {http://arxiv.org/abs/2002.08965}
  {arXiv:2002.08965 [astro-ph.GA]} \BibitemShut {NoStop}%
\bibitem [{\citenamefont {Di~Matteo}\ \emph {et~al.}(2023)\citenamefont
  {Di~Matteo}, \citenamefont {Angles-Alcazar},\ and\ \citenamefont
  {Shankar}}]{DiMatteo:2023jbc}%
  \BibitemOpen
  \bibfield  {author} {\bibinfo {author} {\bibfnamefont {T.}~\bibnamefont
  {Di~Matteo}}, \bibinfo {author} {\bibfnamefont {D.}~\bibnamefont
  {Angles-Alcazar}}, \ and\ \bibinfo {author} {\bibfnamefont {F.}~\bibnamefont
  {Shankar}},\ }\href@noop {} {\  (\bibinfo {year} {2023})},\ \Eprint
  {http://arxiv.org/abs/2304.11541} {arXiv:2304.11541 [astro-ph.HE]}
  \BibitemShut {NoStop}%
\end{thebibliography}%

\newpage
\appendix
\section{SUPPLEMENTAL MATERIAL}
\label{appendix}

\subsection{Distribution of $\Omega_{\rm GW}$ in frequency bins}

Here we give the details of the computation of $P(\Omega)$ in each frequency bin. The total GW energy density \eqref{eq:Omega_i} is given by a sum of contributions $\Omega_{\rm GW}^{(1)}(\vec{\theta}^k_b)$ from individual binaries:
\be
    \Omega_{\rm GW}(f_j)  = \frac{1}{\ln(f_{j+1}/f_j)}\sum_{k=1}^{N(f_j)} \Omega_{{\rm GW},k}^{(1)}  \,,
\ee
where each $\Omega_{{\rm GW},k}^{(1)}$ is independent and identically distributed according to the function $P^{(1)}(\Omega|f_j)$ given in Eq.~\eqref{eq:P1}. As we analyze below the distribution in a single bin, we suppress the argument $f_j$. Although the number of binaries contributing to each bin can be enormous, the distribution does not converge to a Gaussian due to its power-law tail~\cite{Ellis:2023owy}. This tail is generated by a limited number of the strongest events, so we split the GW background into a strong and a weak part:
\be
    \Omega_{\rm GW} = \Omega_{\rm GW,S} + \Omega_{\rm GW,W} \, ,
\ee
where the strong component consists of $N_S$ sources that emit GWs above some threshold $\Omega > \Omega_{\rm th}$ and the remaining sources comprise the weak component. Below we will treat the two components separately and give an analytic estimate for the tail.

\textbf{Strong sources.} The signal from the strong sources is given by 
\be\label{eq:Omega_S}
    \Omega_{\rm GW,S} = \frac{1}{\ln(f_{j+1}/f_j)}\sum_{k=1}^{N_S} \Omega_{\rm GW,S}^{(1)} \,.
\ee 
It is a sum of identical random variables distributed according to
\be\label{eq:P1_S}
    P^{(1)}_{\rm S}(\Omega) = \frac{1}{p_{\rm S}} \, \theta(\Omega - \Omega_{\rm th})P^{(1)}(\Omega)\,,
\ee
where
\be
    p_{\rm S} \equiv \int_{\Omega > \Omega_{\rm th}} \td \Omega \, P^{(1)}(\Omega)
\ee
is the probability of finding a strong source and the expected number of strong sources is
\be
    \bar{N}_{\rm S} = p_{\rm S} \bar{N}\,.
\ee
Note that, since we are interested in a small number of the most dominant sources: $\bar{N}_{\rm S} \ll \bar{N}$, we have $p_{\rm S}\ll 1$. The number of sources $N_S$ in the sum~\eqref{eq:Omega_S} is random and drawn from a Poisson distribution with expectation $\bar{N}_{\rm S}$. The distribution of $\Omega_{\rm GW, S}$ is then given by
\bea\label{eq:P_S_approx}
    P_S(\Omega) = \sum^{\infty}_{N=0} \frac{\bar{N}_{\rm S}^N}{N!}e^{-\bar{N}_{\rm S}} P_S(\Omega|N) \approx P_S(\Omega|\bar{N}_{\rm S}) \,,
\eea
where $P_S(\Omega|N)$ is the distribution of \eqref{eq:Omega_S} with $N$ fixed and the last approximation holds if $\bar{N}_{\rm S} \gg 1$, as in this case the effect of Poisson fluctuations in the number of sources becomes negligible. 

\begin{figure}
    \centering
    \includegraphics[width=0.85\columnwidth]{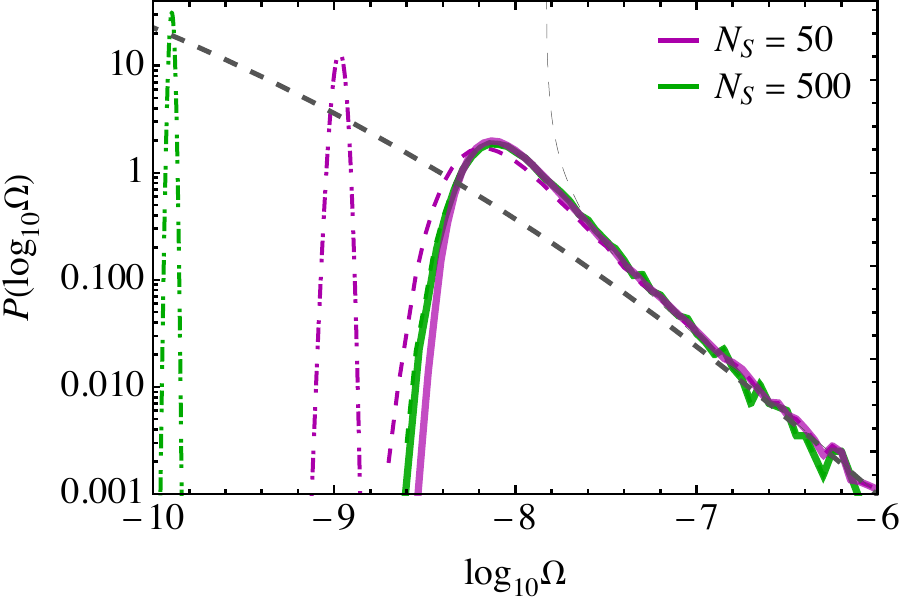}
    \caption{The distribution $P(\log_{10}\Omega|f_j)$ in the fifth NG15 frequency  bin at $f_j = 9.88$ nHz for the best-fit model with environmental effects shown in Fig.~\ref{fig:bins}. The magenta (green) lines show the distributions obtained using $N_{\rm S} = 50$ (500) strong events. The dot-dashed and dashed lines show the distribution of $\log_{10}\Omega$ arising from weak and strong events, respectively. The solid line shows the combined distribution \eqref{eq:P_Omega_tot}. This distribution corresponds to a total of $7 \times 10^7$ individual SMBH binaries with strengths in the range $\log_{10}\Omega^{(1)}_{\rm GW} \in (-25,-3)$. The thick and thin grey dashed line show the analytic estimates~\eqref{eq:tail} and ~\eqref{eq:tail2} of the tail, respectively.}
    \label{fig:P_omega_example}
\end{figure}

The distribution $P_S(\Omega|N)$ can be computed in various ways. In the analysis presented in the main text, we opted for an approach in which we fix $\Omega_{\rm th}$ by demanding $\bar{N}_{\rm S} = 50$. With this choice, $\bar{N}_{\rm S}$ is large enough to apply the approximation in Eq.~\eqref{eq:P_S_approx}, but small enough to allow fast numerical computation of $P_S(\Omega)$: each realization of $\Omega_S$ requires the generation of 50 instances of $\Omega^{(1)}_S$ from the distribution \eqref{eq:P1_S}. To resolve $P_S(\Omega)$, we generate $4\times 10^5$ instances of $\Omega_S$ in each bin. Examples of such distributions with $N_S =50$ and $N_S =500$ are shown by the dashed curves in Fig.~\ref{fig:P_omega_example} . These distributions were generated from $4\times 10^5$ and $4\times 10^4$ random realizations, respectively, corresponding to comparable computation times.

\begin{figure*}
    \centering
    \includegraphics[width=\textwidth]{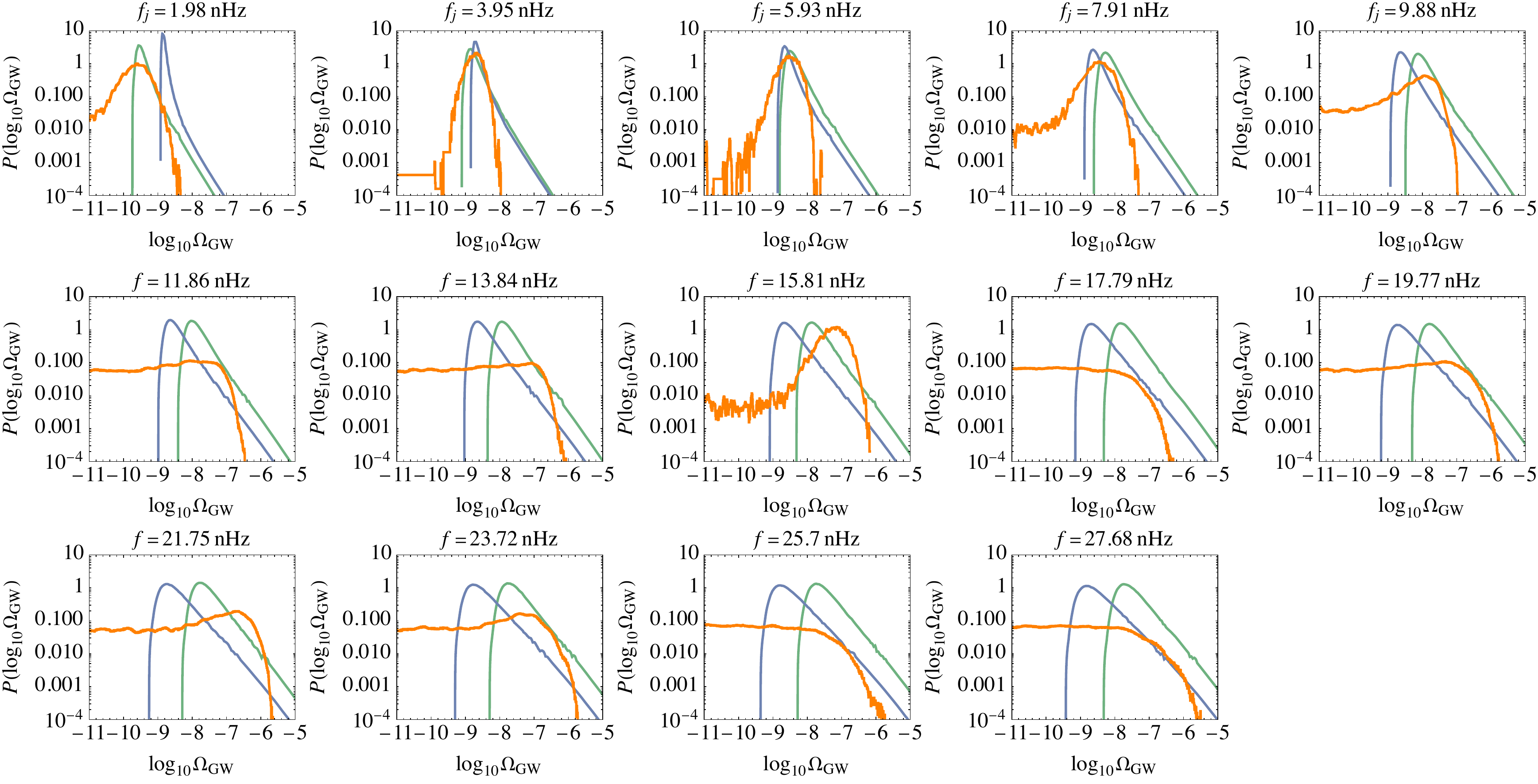}
    \caption{The distributions $P(\log_{10}\Omega|f_j)$ in the 14 NG15 frequency bins for the best fit points with (green) and without (blue) the environmental effects. The orange curves show the NG15 data.}
    \label{fig:bins_2}
\end{figure*}

\textbf{The long tail.} The distribution $P_S(\Omega)$ inherits the long tail from $P^{(1)}(\Omega)$ when $\Omega \to \infty$. We use this to extrapolate $P_S(\Omega)$ to large values of $\Omega$ that are not fully resolved in the Monte Carlo approach. This can be shown by choosing a large enough threshold $\Omega_{\rm th}$ so that $\bar{N}_{\rm S} \ll 1$. Then, only the leading terms in Eq.~\eqref{eq:P_S_approx} contribute, so that 
\be
    P_{\rm S}(\Omega) 
    = (1-\bar{N}_{\rm S})\delta(\Omega) +  \bar{N}_{\rm S} P_{\rm S}(\Omega|1) + \mathcal{O}(\bar{N}_{\rm S}^2) \, ,
\ee
where $P_{\rm S}(\Omega|1) = \ln(f_{j+1}/f_j) P^{(1)}_{\rm S}(\ln(f_{j+1}/f_j) \Omega)$ due to the scaling of the frequency bin width in \eqref{eq:Omega_i} and we used $P_{\rm S}(\Omega|0) = \delta(\Omega)$. Thus, $P_{\rm S}(\Omega) \sim \bar{N}_{\rm S} P_{\rm S}(\Omega|1)$ when $\Omega \to \infty$. The large $\Omega$ tail of the full distribution $P(\Omega)$ is given by $P_{\rm S}(\Omega)$ and does not depend on how the events are split. After the appropriate replacements, we find that
\be\label{eq:tail}
    P(\Omega) \sim \bar{N} \ln(f_{j+1}/f_j) P^{(1)}\!\left[\ln(f_{j+1}/f_j) \Omega\right]\,.
\ee
An improved approximation of the tail can be obtained by shifting the distribution by the expectation value $\bar \Omega$:
\be\label{eq:tail2}
    P(\Omega) \!\!\stackrel{\Omega \gg \bar \Omega}{\approx}\!\! \bar{N} \ln(f_{j+1}/f_j) P^{(1)}\!\left[\ln(f_{j+1}/f_j) (\Omega - \bar \Omega)\right]\!.
\ee
Such a shift preserves the asymptotic behaviour \eqref{eq:tail} and enforces the requirement that the distribution should become strongly peaked around $\bar \Omega$ when the number of sources $\bar{N}$ approaches infinity. This has been derived explicitly in Ref.~\cite{Ellis:2023owy} for an asymptotically $P \sim \Omega^{-5/2}$ distribution using the cumulant generating function. The asymptotic expression \eqref{eq:tail} and the improved approximation~\eqref{eq:tail2} are shown in Fig.~\ref{fig:P_omega_example} by dashed gray lines. Although both match well with the tails obtained via the Monte Carlo approach, the tail~\eqref{eq:tail2} centred around the mean provides a significantly better fit close to the peak.

\textbf{Weak sources.} The total contribution from the weak sources is given by
\be\label{eq:Omega_W}
    \Omega_{\rm GW, W}
    = \frac{1}{\ln(f_{j+1}/f_j)}\sum_{k=1}^{N_{\rm W}} \Omega_{\rm GW,W}^{(1)}
\ee
with the contribution from each source $\Omega_{\rm GW,S}^{(1)}$ distributed as
\be
    P^{(1)}_{\rm W}(\Omega) = \frac{1}{1-p_{\rm S}} \, \theta(\Omega_{\rm th}-\Omega )P^{(1)}(\Omega)\,,
\ee
where the normalization stems from the fact that the probability of a weak source is $1- p_{\rm S} \approx 1$. Analogously, the expected number of weak sources is given by
\be
    \bar{N}_{\rm W} = (1-p_{\rm S}) \bar{N} \approx \bar{N} \gg \bar{N}_{\rm S}\,.
\ee
Since the number of weak sources can be very large ($\bar{N}_{\rm W} \gg 10^3$) and their distribution does not possess a long tail (it is sharply cut at $\Omega > \Omega_{\rm th}$) the sum \eqref{eq:Omega_W} is subject to the central limit theorem and thus $\Omega_{\rm GW,W}$ approximately follows a normal distribution
\be
    P_{\rm W}(\Omega) 
    \approx \frac{1}{\sqrt{2\pi}\sigma_{\rm W}} \exp\!\left[-\frac{(\Omega - \bar{\Omega}_{\rm W})^2 }{2 \sigma_{\rm W}^2} \right]\,,
\ee
where the leading moments are easily found from the moments of a single weak source,
\begin{eqnarray}
    \bar{\Omega}_{\rm W} 
    &=& \frac{\bar{N}_{\rm W}}{\ln(f_{j+1}/f_j)} \left\langle\Omega_{\rm GW,W}^{(1)}\right\rangle\, \nonumber\\
    &\approx& \frac{1}{\ln(f_{j+1}/f_j)}\int_{\substack{\Omega^{(1)}_{\rm GW} \leq \Omega_{\rm th}\\f \in (f_j,f_{j+1})}} \td \lambda \left| \frac{\td t}{\td \ln f_r} \right| \td \ln f \, \Omega^{(1)}_{\rm GW}\,, \nonumber\\
    \sigma_{\rm W} 
    &=& \frac{\sqrt{\bar{N}_{\rm W}}}{\ln(f_{j+1}/f_j)} \sigma^{(1)}_{\rm W}\, ,
%&    = \int_{\substack{\Omega^{(1)}_{\rm GW} \leq \Omega_{\rm th}\\f_r = f_j}}
%    \td \lambda \left| \frac{\td t}{\td \ln f_r} \right| (\Omega^{(1)}_{\rm GW})^2\,,
\end{eqnarray}
with $\sigma^{(1)}_{\rm W} \equiv \sqrt{\langle(\Omega_{\rm GW,W}^{(1)})^2\rangle - \langle\Omega_{\rm GW,W}^{(1)}\rangle^2}$. An instance of the Gaussian approximation for the weak sources is shown by the dot-dashed lines in Fig.~\ref{fig:P_omega_example} for $N_S =50$ and $N_S =500$.

\textbf{The distribution of $\Omega_{\rm GW}$.} Combining the strong and weak sources gives
\be\label{eq:P_Omega_tot}
    P(\Omega) 
    = \int \td \Omega' P_{\rm W}(\Omega - \Omega')  P_{\rm S}(\Omega')
    \approx P_{\rm S}(\Omega - \bar{\Omega}_{\rm W}) \, ,
\ee
where the last approximation holds when $P_{\rm W}$ is significantly narrower than $P_{\rm S}$. As shown in Fig.~\ref{fig:P_omega_example}, the combined distribution works well for $\bar{N}_{\rm S} = 50$. The differences from the $\bar{N}_{\rm S} = 500$ case are negligible and appear in the small $\Omega$ tail. They can be ascribed to Poisson fluctuations in the MC method as well as neglecting the fluctuations in the number of contributing events centred around $\bar{N}_{\rm S}$. Also, when $\bar{N}_{\rm S} = 500$, the contribution of weak sources becomes negligible and the strong sources alone can account for the distribution of the signal.

The explicit distributions $P(\log_{10}\Omega|f_j)$ in all 14 NG15 frequency bins for the best-fit scenarios with and without environmental effects are collected in Fig.~\ref{fig:bins_2} and compared to the NG15 data. It contains essentially the same information as Fig.~\ref{fig:bins}, but one can clearly see power-law tails that characterise sizeable upward fluctuations.

\begin{figure}[b]
    \centering
    \includegraphics[width=0.84\columnwidth]{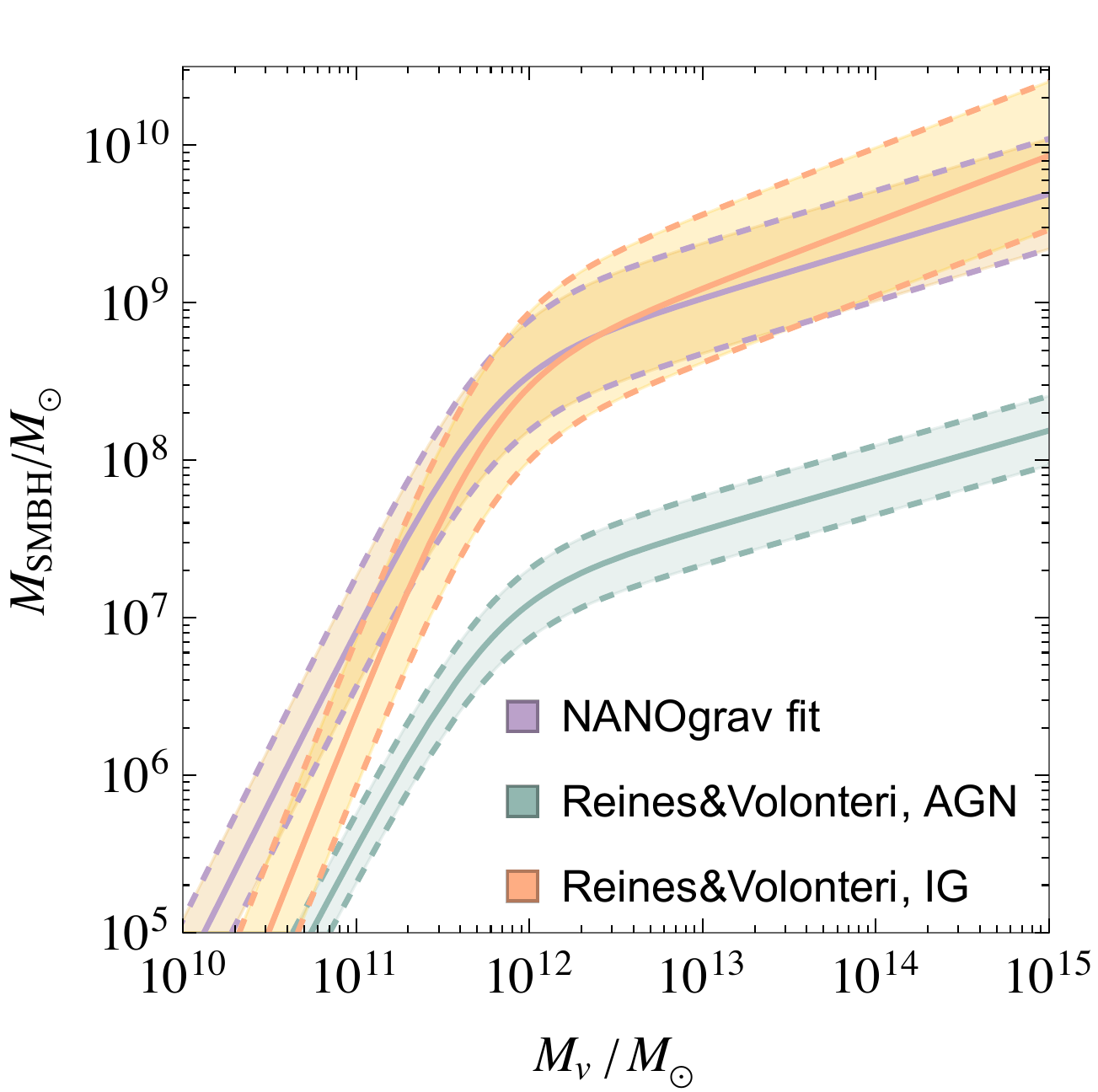}
    \caption{The green and orange curves show the local SMBH mass-halo mass relations arising from the SMBH mass-stellar mass relations for AGNs and IGs from~\cite{2015ApJ...813...82R} and the stellar mass-halo mass relation from~\cite{Girelli:2020goz}. The purple curve shows the phenomenological fit from the NANOGrav Collaboration~\cite{NANOGrav:2023hfp}. The solid curves show the best fits and the band width indicates the intrinsic scatter.}
    \label{fig:scaling_plot}
\end{figure}

\subsection{SMBH mass-halo mass relation}

The basis of the relationship between the SMBHs and their host galaxies appears to be through the properties of the bulge and observed correlations that hint at their coevolution~\cite{Kormendy:2013dxa}. However, there is a sizeable difference in the SMBH mass-host mass relation, around an order of magnitude, in local active (AGN) and inactive (IG) galaxies~\cite{2015ApJ...813...82R}. Since we rely on the EPS merger rate of halos, we have translated the stellar mass to halo mass using the fits provided in Ref.~\cite{Girelli:2020goz}. In Fig.~\ref{fig:scaling_plot} we show the local ($z=0$) SMBH mass-host mass relation obtained from the SMBH mass-stellar mass relation to both AGNs and IGs~\cite{2015ApJ...813...82R}. For comparison, we show also the phenomenological fit found by the NANOGrav Collaboration using the NG15 data~\cite{NANOGrav:2023hfp} and priors from~\cite{Kormendy:2013dxa}. In our main analysis, we use the IG mass relation and its scatter, since only a small fraction of the local galaxies are active.

\begin{figure}
    \centering
    \includegraphics[width=0.9\columnwidth]{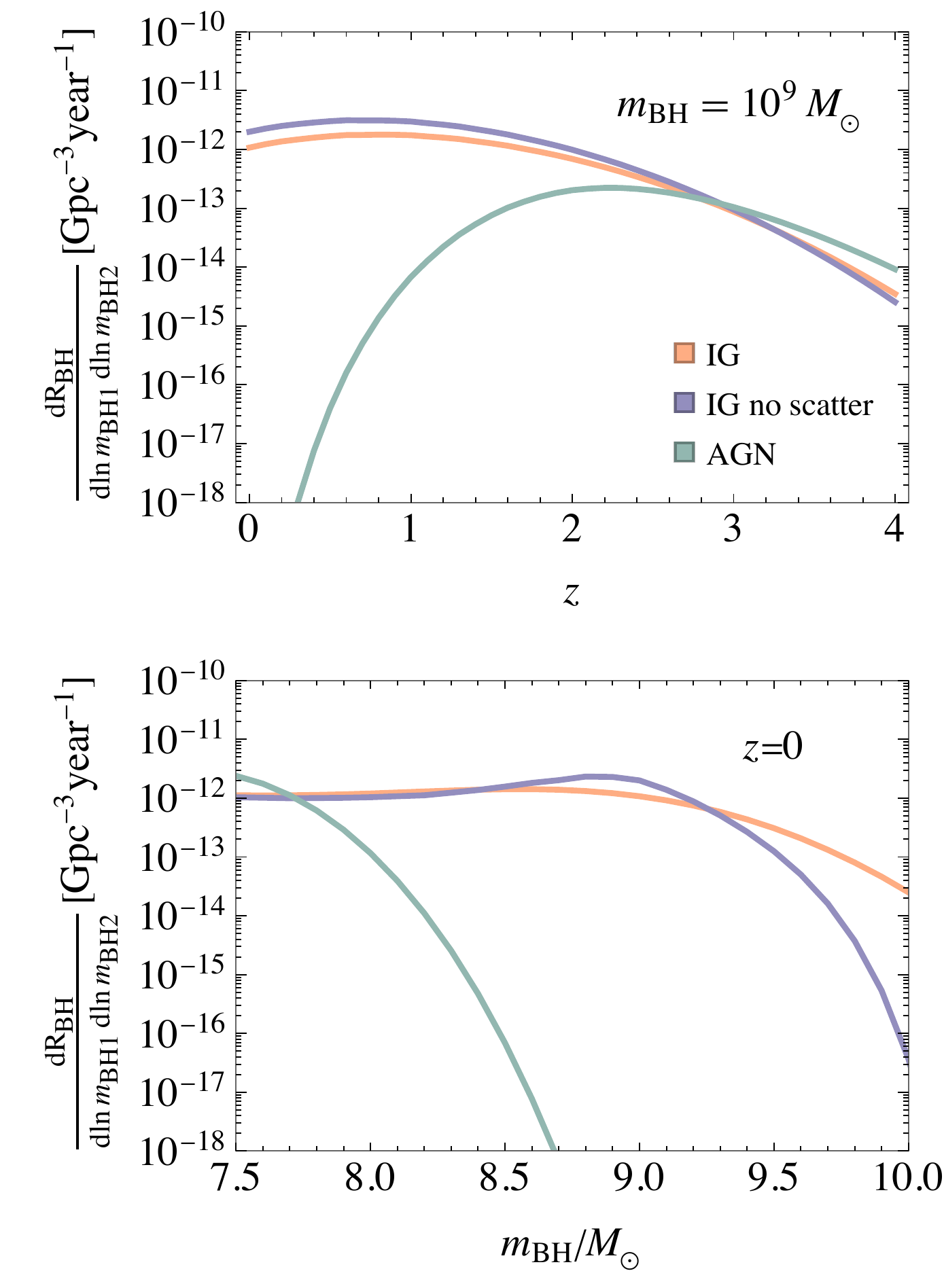}
    \vspace{-5mm}
    \caption{Merger rates of equal mass SMBH binaries for the mass relations shown in Fig.~\ref{fig:scaling_plot} for IGs with and without scatter and for AGNs.}
    \label{fig:mr_plot}
\end{figure}

Recent analyses of the JWST data~\cite{2023arXiv230311946H,2023ApJ...954L...4K,Maiolino:2023bpi} have found AGNs in galaxies at $z=4-7$ that are heavier than in the local galaxies, and Ref.~\cite{Pacucci:2023oci} showed that the high-$z$ mass relation deviates by more than $3\sigma$ from the local relation. Their results further indicate that only the overall prefactor of the mass relation changes while the tilt and scatter remain roughly intact. Therefore, for AGNs, we extend the relation given in Ref.~\cite{2015ApJ...813...82R} with a power-law $z$ dependence $\log_{10}(\bar{m}/M_\odot) = 7.45 + 1.05 \log_{10} (M_*/10^{11}M_\odot) + 2.35 \log_{10} (1+z)$. 

The high-$z$ mass relation measured from AGNs is quite similar to that measured in local IGs. Given that the fraction of active galaxies grows with $z$~\cite{Aversa:2015bya,2020ApJ...892...17D}, this suggests that SMBH masses increase as fast as the stellar masses and that the present-day AGNs reflect the low-mass tail of the SMBH mass distribution, motivating our choice not to include $z$ dependence in the SMBH mass-stellar mass relation.

\begin{figure}
    \centering
    \includegraphics[width = 0.9\columnwidth]{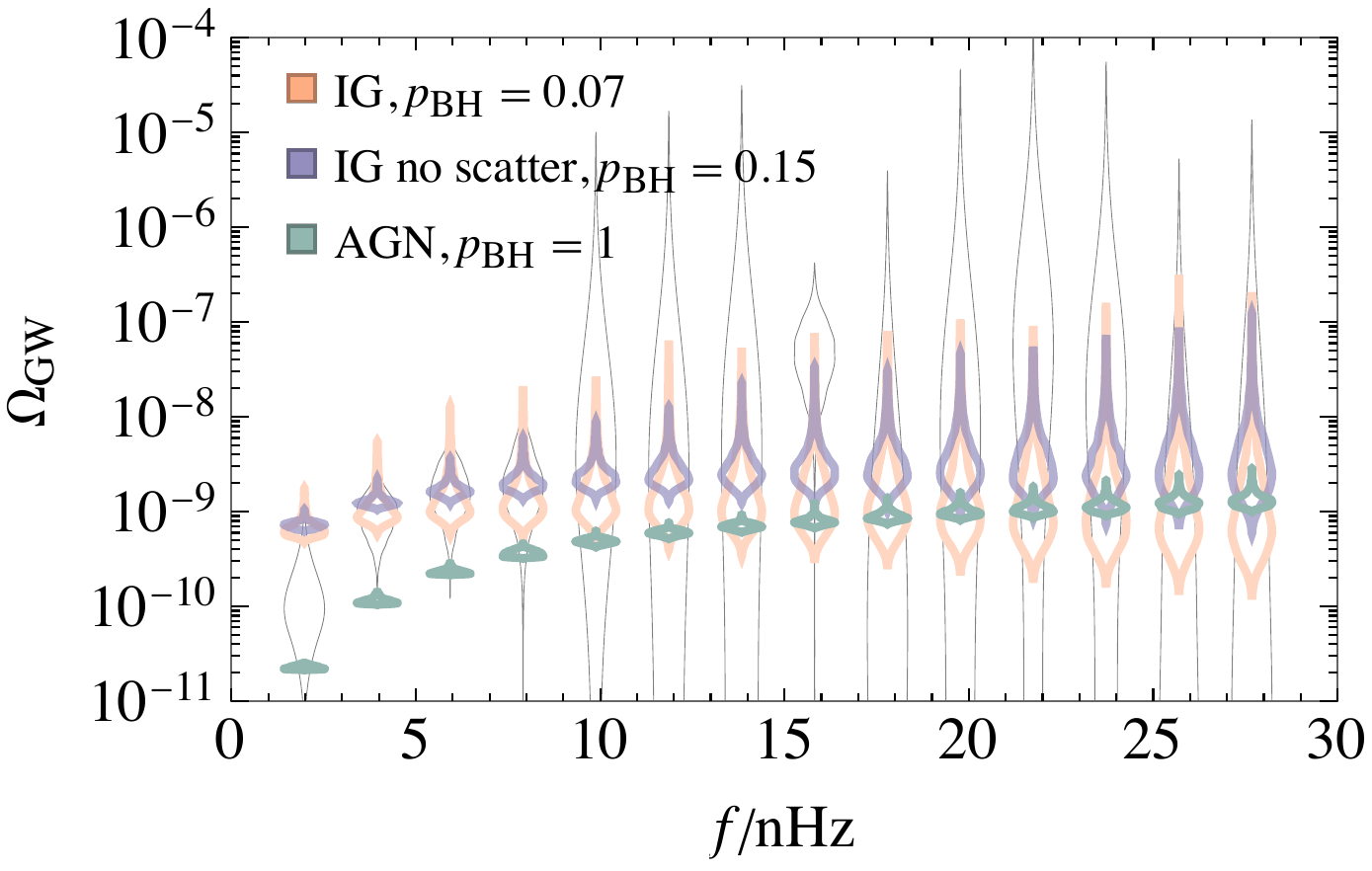}
    \vspace{-2mm}
    \caption{The best fits to the NG15 data (thin lines) generated by SMBH with GW-only evolution, with and without the scatter in the SMBH mass-stellar mass relation for IGs and for the AGNs.}
    \label{fig:no_spread}
\end{figure}

As can be seen in Fig.~\ref{fig:mr_plot}, changes in the mass relation can significantly affect the SMBH merger rate, especially at the heavy end of the mass spectrum ($m_{\rm BH}>10^8 M_{\odot}$) and at low redshifts ($z<2$). The exponential drop in the rate as a function of mass reflects the behavior of the halo mass function. The rate using the mass relation for AGNs falls off quickly at high masses at low-$z$ because, as can be seen from Fig.~\ref{fig:scaling_plot}, the heavy local AGN SMBHs are in much heavier halos than heavy local SMBHs in inactive galaxies.

Fig.~\ref{fig:no_spread} shows the best fits to the NG15 data in the GW-only driven case for the rates shown in Fig.~\ref{fig:mr_plot}. The AGN-only fit can barely reproduce the observed background for $p_{\rm BH} = 1$ and decreases the goodness of the fit with $\Delta \chi^2 = 43$ compared to the IG fit. This indicated that there must be a significant bias in the detection of AGN or that they represent just the lightest cue of the population. This possibility is further supported by the results from numerical cosmological simulations which prefer SMBH in the heavy side of the observations, see \cite{DiMatteo:2023jbc} for a review. One effect of suppressing the rate of the heaviest binaries is that the distribution of $\Omega_{\rm GW}$ becomes significantly narrower. Interestingly, this also demonstrates that information about the SMBH mass function can be extracted from measurements of fluctuations in the signal between frequency bins.

Another effect we have explored is that of the scatter in the SMBH-host mass relations. We have performed the analysis using the IG mass relation from~\cite{2015ApJ...813...82R} without accounting for the scatter. The resulting rate is shown in purple in Fig.~\ref{fig:mr_plot} and the fit in purple in Fig.~\ref{fig:no_spread}. The best fit is shifted to a slightly bigger value of $p_{\rm BH}=0.15$ compared to the fit with the scatter ($p_{\rm BH}=0.07$), and the spectrum changes such that the amplitude $\Omega_{\rm GW}$ decreases faster with frequency. Neglecting scatter in the mass relation worsens the best fit by $\Delta \chi^2 = 1.53$. In all, the effect of the scatter is not very significant. 
\end{document}